\documentclass[proof]{WileyASNA-v1}

\newcommand{\arcsec}{''}
\newcommand{\degr}{$^\circ$}
\newcommand{\kms}{km\,s$^{-1}$}
\newcommand{\Halpha}{H$\alpha$}
\newcommand{\Hbeta}{H$\beta$}
\newcommand{\xiA}{$\xi$\,Boo\,A}
\newcommand{\xiB}{$\xi$\,Boo\,B}
\newcommand{\beq}{\begin{equation}}
\newcommand{\eeq}{\end{equation}}
\newcommand{\Teff}{$T_{\rm eff}$}
\newcommand{\liseven}{$\mathrm{^{7}Li}$}
\newcommand{\lisix}{$\mathrm{^{6}Li}$}
\newcommand\ion[2]{#1$\;${\scshape{#2}}}

\articletype{Original Paper}

\received{<day> <Month>, <year>}
\revised{<day> <Month>, <year>}
\accepted{<day> <Month>, <year>}

\begin{document}

\title{On the lithium abundance of the visual binary components $\xi$~Boo~A (G8V) and $\xi$~Boo~B (K5V)}

\author[1]{Klaus G. Strassmeier$^1$*}

\author[1]{Matthias Steffen$^1$*}

\authormark{K. G. Strassmeier \& M. Steffen}

\address[1]{\orgname{Leibniz-Institute for Astrophysics Potsdam (AIP)}, \orgaddress{An der Sternwarte 16, D-14482 Potsdam, \country{Germany}}}

\corres{*K. G. Strassmeier, AIP \email{kstrassmeier@aip.de}\\M. Steffen, AIP \email{msteffen@aip.de}}

\abstract[Abstract]{
A spectroscopic investigation of the lithium resonance doublet in \xiA\ and \xiB\ in terms of both abundance and isotopic ratio is presented. We obtained new $R$=130\,000 spectra with a signal-to-noise ratio (S/N) per pixel of up to 3200 using the 11.8m LBT and PEPSI. From fits with synthetic line profiles based on 1D-LTE MARCS model atmospheres and 3D-NLTE corrections, we determine the abundances of both isotopes. For \xiA, we find A(Li) = 2.40$\pm$0.03\,dex and $^6$Li/$^7$Li < 1.5$\pm$1.0\,\%\ in 1D-LTE, which increases to $\approx$2.45 for the 3D-NLTE case. For \xiB\ we obtain A(Li) = 0.37$\pm$0.09\,dex in 1D-LTE with an unspecified $^6$Li/$^7$Li level. Therefore, no $^6$Li is seen on any of the two stars. We consider a spot model for the Li fit for \xiB\ and find A(Li) = 0.45$\pm$0.09\,dex. The $^7$Li abundance is 23 times higher for \xiA\ than the Sun's, but three times lower than the Sun's for \xiB\ while both fit the trend of single stars in the similar-aged M35 open cluster. Effective temperatures are redetermined from the TiO band head strength. We note that the best-fit global metallicities are --0.13$\pm$0.01\,dex for \xiA\ but  +0.13$\pm$0.02\,dex for \xiB. Lithium abundance for the K5V benchmark star 61\,Cyg\,A was obtained to A(Li)$\approx$0.53\,dex when including a spot model but to $\approx$0.15\,dex without a spot model.
}


\maketitle

\footnotetext{\textbf{Abbreviations:} PEPSI Potsdam Echelle Polarimetric and Spectroscopic Instrument. LBT Large Binocular Telescope. (N)LTE (Non)Local Thermodynamic Equilibrium. }

\section{Introduction}\label{S1}

$\xi$~Boo (HIP\,72659, GJ566) is a nearby visual binary with two bright components denoted as {$\xi$\,Boo\,A} (HD\,131156A) and {$\xi$\,Boo\,B} (HD\,131156B). The AB orbital period is 151\,yr in a strongly inclined 0.51-eccentricity orbit (Wielen 1962). Its inclination of 140\degr\ means retrograde motion with respect to increasing position angles and a residual inclination with respect to the plane of the sky of 50\degr . The current apparent AB separation is 7.15\arcsec . The A primary component is a solar-like star of spectral type around G8V while the B secondary component is a significantly cooler and lower mass star of spectral type K4-5V (Abt 1981, Levato \& Abt 1978). Both components have moderately strong and variable Ca\,{\sc ii} emission (Wilson 1978) with irregular long-term fluctuations (Lockwood et al. 2007, Baliunas et al. 1995). Its strengths indicate surface magnetic activity exceeding that of the Sun by about a factor 2--3. Due to its brightness the system had been the target of numerous observations ranging from X-rays (e.g., Johnstone \& G\"udel 2015; Wood et al. 2018) to radio wavelengths (e.g., Linsky \& Gary 1983 and references therein).

\begin{figure*}[!h!tb]
\centering
\includegraphics[clip,angle=0,width=\textwidth]{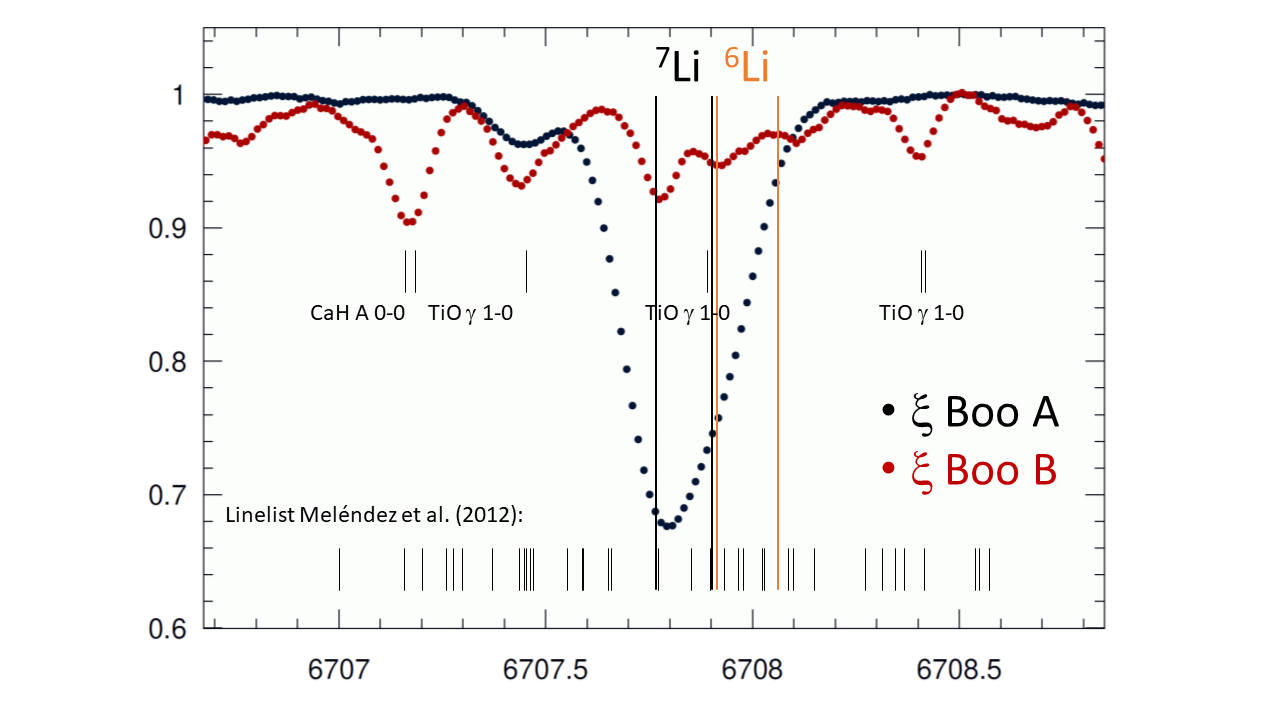}
\caption{Lithium 6708\,\AA\ of \xiA\ (black dots) and \xiB\ (red dots). The wavelengths of the two doublets from the two lithium isotopes are marked as vertical lines. Also indicated are the blending features from the line list of Mel\'endez et al. (2012). Additional molecular features from the sunspot umbral spectrum of Wallace et al. (1999) are indicated as well. }\label{F1}
\end{figure*}

Of most interest in the context of the present paper is the fact that the A
component shows strong Li\,{\sc i} 6708-\AA\ absorption while the B component
has an essentially undetected Li\,{\sc i} 6708-\AA\ line. This was already
noticed by Wilson (1963) and Herbig (1965), and confirmed by Savanov
(1992). The absent Li in the B component had been attributed to the expected
strong convective mixing and thermonuclear destruction of Li in such a very
cool star. The presence of strong Li in the G8V component with an abundance of
A(Li)\footnote{A(Li) = $\log(N({\rm Li})/N({\rm H})) + 12$.}$\approx$2.4 (Luck
2017) is a sign of youth, or the lack of above mentioned convective mixing. It
appears not an uncommon value among late-to-mid G stars only slightly older
than the Pleiades for which Bouvier et al. (2018) quote abundances of
$\approx$2.7 for temperatures around 5500\,K, but noticed and emphasized the
large spread of 0.5--2.5 for the cooler temperatures around $\approx$4600\,K.

\begin{table*}[htb]
\caption{Adopted astrophysical properties of \xiA\ and \xiB . } \label{T1}
\begin{center}
\begin{tabular}{lllll}
\hline \noalign{\smallskip}
Parameter                   & \xiA  & Ref. & \xiB   & Ref.  \\
\noalign{\smallskip}\hline \noalign{\smallskip}
Classification, MK          & G8V & (1) & K5V & (1) \\
Effective temperature, K    & 5480 & (9) & 4570 & this paper \\
Log gravity, cm\,s$^{-2}$   & 4.53 & (9) & 5.0 & (9)\\
$v\sin i$, \kms             & 3.0 & (2) & 1.5 & this paper \\
Microturbulence, \kms       & 1.40 & (9) & 0.15 & (9) \\
Macroturbulence, \kms       & 3.6 & (3) & \dots &  \\
Rotation period, d          & 6.43 & (6) & 11.94 & (8) \\
Inclination, deg            & 28$\pm$5 & (5,11) & $\approx$32 & this paper\\
Iron abundance, H=12        & 7.29 & (9) & 7.45 & (9) \\
Distance, pc                & 6.753 & (10) & 6.748 & (10) \\
Luminosity, L$_\odot$       & 0.55 & (9) & 0.17 & this paper \\
Radius, R$_\odot$           & 0.82 & (9,10) & 0.66 & assumed \\
Age, Myr                    & 187 & (12) & 265 & (12) \\
\noalign{\smallskip}\hline
\end{tabular}
\end{center}
\vspace*{3mm}
\begin{tablenotes}
{(1) Abt (1981). (2) Gray (1994). (3) Allende Prieto et al. 2004. (4) Fernandes et al. (1998). (5) Petit et al. (2005). (6) Toner \& Gray (1988). (7) Ruck \& Smith (1995). (8) Donahue et al. (1996). (9) Luck (2017). (10) {\it Gaia} DR-3 (2022). (11) Morgenthaler et al. (2012). (12) Barnes (2007). }\\
\end{tablenotes}
\end{table*}

Standard stellar evolution models predict already significant lithium depletion during the pre-main sequence phase, whereas they wrongly predict little or no depletion during the main sequence (e.g., Iben 1965). Therefore, the standard models overestimate the present-day solar lithium abundance by about a factor of 100, and thus are also at odds with the depletion pattern observed in open clusters (e.g., Bouvier et al. 2018). With $\xi$~Boo AB, we have likely two coeval stars with rather different convective envelopes and thus mixing efficiencies which allows to place some more tighter limits on the Li evolution. Regarding the importance of using one-dimensional (1D) vs. three-dimensional (3D) atmospheres in LTE vs. non-LTE (NLTE), we refer to our previous papers by Mott et al. (2020, 2017) and Harutyunyan et al. (2018).

Section~\ref{S2} describes our new observations. Section~\ref{S3} is a review of the stellar parameters of $\xi$ Boo. Section~\ref{S4} is the analysis for the two stars including 61\,Cyg\,A for comparison, and Section~\ref{S5} a brief summary and conclusion. In the Appendix, we address the question whether star spots are the cause of enhanced TiO line strength and present a numerical study of TiO line strengths as a function of stellar effective temperature, star spot temperature and area filling factor.

\section{Observations}\label{S2}

New high-resolution, high S/N spectra were obtained with PEPSI (Strassmeier et
al. 2015) at the effective 11.8\,m LBT (Hill et al. 2012) in southern
Arizona. The two 8.4\,m LBT mirrors (dubbed SX and DX) were used in binocular
mode, that is like a single 11.8\,m telescope, which for PEPSI is achieved by
combining the spectra from the two mirrors after integration. For the current
observations PEPSI itself was fed through its IQUV polarimeters at the two
straight-through Gregorian foci. For Stokes~I all six individual QUV exposures
were combined for its final spectrum according to a total exposure time of
30\,min for \xiA\ and 60\,min for \xiB . The two pairs of 200\,$\mu$m fibers
result in a spectral resolution of $R=\lambda/\Delta\lambda$=130\,000 sampled
by 4.2 pixels.

Observations of both $\xi$\,Boo components were spread over 10 consecutive
nights in May 6-16, 2019. Eight and six individual spectra for \xiA\ and
\xiB\, respectively, were obtained with cross disperser (CD) III covering
4800--5441~\AA\ and with CD~V covering 6278--7419~\AA , the latter contains
the lithium line at 6708\,\AA . Six individual exposures with an exposure time
of 5\,min for \xiA\ and 10\,min for \xiB\ make up one phase spectrum. It
resulted in S/N per pixel of up to 3200 in CD~V with an average of 2,670 at
the location of the lithium line for \xiA , and up to 1700 for \xiB\ with an
average of 1450 at 6708\,\AA. Note that one spectrum of \xiB\ was taken during
bad weather conditions and reached S/N of just 215. Fig.~\ref{F1} shows
example spectra for both stellar components. The log of all observations is
given in the Appendix in Table~\ref{T1-App}.

Data reduction was performed semi-automatically with the software package
SDS4PEPSI (Spectroscopic Data Systems for PEPSI) based on the original code of
Ilyin (2000), and described in some detail in Strassmeier et al. (2015,
2018a). The specific steps of image processing include bias subtraction and
variance estimation of the source images, super-master flat field correction
for the CCD spatial noise, scattered light subtraction, definition of
\'echelle orders, wavelength solution for the ThAr images, optimal extraction
of image slicers and cosmic spikes elimination, normalization to the master
flat field spectrum to remove CCD fringes and the blaze function, a global 2D
fit to the continuum, and the rectification of all spectral orders into a 1D
spectrum.

\section{Review of relevant $\xi$\,Boo~A and $\xi$\,Boo~B data}\label{S3}

Table~\ref{T1} summarizes the stellar input parameters for both components for
our lithium fit. \xiA's fundamental stellar parameters were collected and
discussed in Petit et al. (2005). This paper serves as the basis for our
updated and extended summary. Note that a total of 29 papers are listed in
CDS/Simbad for \xiA\ which present a determination of effective temperature
$T_{\rm eff}$, gravity $\log g$, and/or metallicity [M/H] or relative iron
abundance [Fe/H]. A mean effective temperature of 5550\,K for \xiA\ was
obtained by Gray (1994) from spectral line ratios. Savanov (1992) applied a
model-atmosphere analysis to their high-dispersion photographic spectra and
obtained effective temperatures for A and B of 5300$\pm$100\,K and
$\approx$4300\,K, and gravities of 4.1$\pm$0.2 and $\approx$4.5,
respectively. The metallicity of the primary was initially given as $-0.20\pm
0.08$ by Cayrel de Strobel et al. (1992). Ruck \& Smith (1995) presented a
fine analysis of calcium and iron lines from $R$=100\,000 spectra of \xiA\ and
found 5500$\pm$70\,K, $\log g$=4.6$\pm$0.1, and relative abundances of [Fe/H]
and [Ca/H] of $-0.15$ and $-0.13$, respectively, with typical errors of
$\pm$0.05. The S4N catalog (Allende Prieto et al. 2004) lists component A with
5350$\pm$115\,K, gravity of 4.576$\pm$0.050, and an absolute abundance
[Fe/H]=7.33, obtained with a microturbulence of 1.19\,\kms\ and a (Gaussian)
macroturbulence of 3.64\,\kms. The analysis in the catalog of Luck (2017)
provides probably the most consistent stellar parameters. They list \xiA\ with
$T_{\rm eff}$=5480$\pm$33\,K, $\log g$=4.53, with a microturbulence of
1.38\,\kms, a $v\sin i$ of 6.0\,\kms, and an absolute [Fe/H] abundance of
7.29$\pm$0.05. No macroturbulence is stated though, which may explain the
discrepancy in $v\sin i$ to other measurements, for example those from Gray
(1994) or Toner \& LaBonte (1991) of $\approx$3\,\kms.

For component B, Luck (2017) derived $T_{\rm eff}$ of 4767\,K, $\log g$ of
5.0, with a microturbulence of 0.15\,\kms, a $v\sin i$ of 5.1\,\kms, and an
absolute [Fe/H] abundance of 7.45$\pm$0.15 (no other errors given). Our
spectra show that \Halpha\ and \Hbeta\ appear as wingless but otherwise strong
absorption lines. The rotational line broadening is a factor two smaller for
\xiB\ than for \xiA , in particular for temperature and gravity insensitive
lines, while line equivalent widths for \xiB\ can be larger by a factor 2--3
for gravity-sensitive lines compared to \xiA. This may have been reflected in
the \xiB\ analysis of Luck (2017) by their comparably small microturbulence
broadening of 0.15\,\kms\ combined with a too large $v\sin i$ of
5.1\,\kms. Takeda et al. (2007) used spectra from the Keck-based SPOCS catalog
(Valenti \& Fischer 2005) to also obtain absolute parameters for both
components. While their results for \xiA\ ($T_{\rm eff}$=5570$\pm$31\,K, $\log
g$=4.57$\pm$0.02, [Fe/H]=--0.07$\pm$0.02) are consistent with other
determinations, their values for \xiB\ have a lower gravity (4.40) and thus a
higher mass (0.99\,M$_\odot$) than for component A (0.93\,M$_\odot$), and thus
appear to be grossly inconsistent with the component's apparent
brightness. The {\it Gaia} DR-3 parallaxes of both components are practically
identical (6.753$\pm$0.006\,pc for A and 6.748$\pm$0.002 for B) and, together
with the fact that they form a physically connected binary, makes them likely
also coeval. We also note that the DR3-based Apsis FLAME luminosity and radii
for both components (A: 0.55$\pm$0.01 L$_\odot$, 0.86$\pm$0.02 R$_\odot$; B:
0.13$\pm$0.01 L$_\odot$, 0.66$\pm$0.02 R$_\odot$) agree very well with our
values in Table~\ref{T1}.

Age determinations for the two components are unsurprisingly widespread. Most
of the values come from comparisons of the spectroscopically determined
fundamental parameters with evolutionary tracks and their
isochrones. Fernandez et al. (1998) were among the first with a consistent
modern analysis that led them to ages of 2$\pm$2\,Gyr, already indicating the
complexity of the data situation. Takeda et al. (2007) compared their
SPOCS-based stellar parameters with YREC tracks and derived the rather
inconsistent ages of $<$0.76\,Gyr for \xiA\ and 12.60\,Gyr for \xiB. Based on
Ca\,{\sc ii} H\&K emission and a common age-activity relation, Wright et
al. (2004) obtained ages of $0.00-0.35$\,Gyr for A and $0.18-3.89$\,Gyr for
B. The measured rotation periods of A and B of 6.4 and 11.9\,d, together with
mean $B-V$'s of 0.76\,mag and 1.17\,mag for the two components, respectively,
led Barnes (2007) to a 187\,Myr based gyrochronological age for component A
and 265\,Myr for component B. Given the inherent uncertainties in
gyrochronology, these ages are likely consistent with each other. No
asteroseismic ages for either component are available to date.


\section{Analysis}\label{S4}

\subsection{Model atmospheres and spectrum synthesis}

We basically follow our previous Li analysis in Mott et al. (2017) updated by the results from the detailed 3D non-LTE vs. 1D LTE comparisons in Mott et al. (2020) and Harutyunan et al. (2018). Therefore, in the present paper, we only synthesize 1D-LTE spectra for the astrophysical parameters of \xiA\ and \xiB\ and then apply 3D non-LTE corrections if available from previous calibrations. The Turbospectrum package (Plez 2012) is employed under the assumption of LTE to create the synthetic spectra.

The model atmospheres we used are those from MARCS (Gustafsson et al. 2008). A sample of atmospheres is chosen such that they bracket the stellar parameters listed in Table~\ref{T1}. A total of 400 spectra are synthesized per model atmosphere covering 20 Li abundances and five isotope ratios (each for four values for the microturbulence) and thus cover a 6-dimensional parameter space including $T_{\rm eff}$, $\log g$, [Fe/H], and microturbulence $\xi_{\rm micro}$. The latter is not used as a free parameter in the line fit but was fixed to the values given in Table~\ref{T1}. Note that it describes a depth-independent isotropic Gaussian velocity distribution with a dispersion of $v_{\rm rms} = \xi_{\rm micro}/\sqrt{2}$.

\begin{table*}[!tbh]
\caption{1D-LTE Li results for \xiA\ and \xiB . } \label{T2}
\center
\begin{tabular}{lllllll}
\hline\noalign{\smallskip}
 BJD mid      & $\phi$ & S/N & [M/H] & $^6$Li/$^7$Li  &  A(Li)   & $\chi^2$ \\
 (+2450000)   &        &     & solar  & (\%)           &  (H=12)  & fit\\
\noalign{\smallskip}\hline\noalign{\smallskip}
\xiA : & & & & & & \\
 8609.7427427 & 0.829 &  2861& --0.126$\pm$0.0015 & 1.448$\pm$0.001 & 2.400$\pm$0.001 & 1143 \\
 8610.8282148 & 0.998 &  2736& --0.127$\pm$0.0016 & 1.692$\pm$0.001 & 2.400$\pm$0.001 & 1148 \\
 8611.7681903 & 0.144 &  2228& --0.130$\pm$0.0020 & 1.872$\pm$0.001 & 2.398$\pm$0.001 & 791 \\
 8616.8333173 & 0.932 &  1688& --0.130$\pm$0.0027 & 1.381$\pm$0.001 & 2.402$\pm$0.001 & 535 \\
 8617.6923984 & 0.066 &  2731& --0.130$\pm$0.0016 & 1.828$\pm$0.001 & 2.400$\pm$0.001 & 1166 \\
 8617.8769492 & 0.094 &  3233& --0.135$\pm$0.0014 & 1.779$\pm$0.001 & 2.399$\pm$0.001 & 1833 \\
 8619.7004465 & 0.378 &  2666& --0.126$\pm$0.0017 & 1.145$\pm$0.001 & 2.399$\pm$0.001 & 1327 \\
 8619.8758547 & 0.405 &  3195& --0.131$\pm$0.0014 & 1.283$\pm$0.001 & 2.399$\pm$0.001 & 1744 \\
 {\bf Average}&       &      & {\bf --0.129$\pm$0.0017} & {\bf 1.554$\pm$0.001} & {\bf 2.400$\pm$0.001} &  \\
 \noalign{\smallskip}
 \xiB : & & & & & & \\
 8609.7860453 & 0.124 & 1713& +0.127$\pm$0.002 & \dots & 0.367$\pm$0.001 & 5524 \\
 8610.8816143 & 0.215 &  215& +0.138$\pm$0.014 & \dots & 0.47$\pm$0.01  & 85 \\
 8611.8163769 & 0.294 & 1560& +0.131$\pm$0.002 & \dots & 0.366$\pm$0.001 & 4465\\
 8616.8730350 & 0.717 & 1569& +0.129$\pm$0.002 & \dots & 0.369$\pm$0.001 & 4514 \\
 8617.8435912 & 0.798 & 1366& +0.127$\pm$0.002 & \dots & 0.368$\pm$0.001 & 3564 \\
 8619.7390032 & 0.957 & 1015& +0.127$\pm$0.003 & \dots & 0.362$\pm$0.002 & 1924\\
 {\bf Average$^a$}      &       &     & {\bf +0.128$\pm$0.002} & \dots & {\bf 0.366$\pm$0.003} &  \\
 \noalign{\smallskip}\hline
\end{tabular}
\begin{tablenotes}
BJD is barycentric-coordinate Julian date for the time of mid exposure. $\phi$ is the rotational phase based on the respective ephemeris for \xiA : 2,452,817.41  +  6.43 $\times\ E$, and for \xiB : 2,452,817.41  +  11.94 $\times\ E $. S/N is given for the pixel at the continuum near 6708.5\,\AA. Errors are always internal errors and were less than $10^{-3}$ for A(Li) for \xiA\ but were round up to $10^{-3}$ in the table. The $\chi^2$ value of the fit refers to a 130-pixels and 72-pixels range in wavelength space centered at Li\,{\sc i} for \xiA\ and \xiB , respectively. $^a$without the value at fractional BJD 8610.88.
\end{tablenotes}
\end{table*}

\subsection{Line list}

Based on the four line-list comparisons in a previous application by Mott et al. (2017), we favor the line list collected by Mel\'endez et al. (2012) expanded by the vanadium revision of Lawler et al. (2014). Lawler et al. (2014) provided improved values of both wavelength and oscillator strength for the V\,{\sc i} blend close to one of the $^6$Li components ($\lambda$ = 6708.1096\,\AA, $\log gf$ = $-2.63$). We use these values for V\,{\sc i} 6708.094\,\AA\ instead of the ones in Mel\'endez et al. (2012). Apart from lithium, the total number of lines in this list is 36. The wavelengths are indicated as short vertical dashes in Fig.~\ref{F1}. The four lithium transitions are implemented with their hyper-fine structure (HFS) with a total of 12 line components and are based on Kurucz (2006). For each isotope separately, we add up the hyperfine fractional strengths of transitions between the same fine structure levels (characterized by quantum number $J$) but different hyperfine levels (characterized by quantum number $F$) that have (nearly) identical wavelengths. In this way, ten transitions of \liseven\ can be reduced to six, and nine transitions of \lisix\ can also be reduced to six, resulting in the 12-component representation of the isotopic HFS of the lithium resonance doublet given originally in Mott et al. (2017). The main differences between Kurucz (1995), his  Table~1, and Kurucz (2006) are slight wavelength shifts of the HFS components of up to 2~m\AA\ for \lisix\ and up to 8~m\AA\ for \liseven. Wavelength uncertainties of this order are considered irrelevant since such changes in the wavelength shifts are a factor 20 smaller than the separation of the fine structure doublet components of each isotope.

The line data of other references are less detailed than those of Kurucz.  For example, Andersen et al. (1984) lists only four components of \liseven\ and two components of \lisix. Smith et al. (1998) and Hobbs et al. (1999) as well as Mel\'endez et al. (2012) give wavelengths and $\log gf$ values for six components of \liseven\ and three components of \lisix. For the purpose of comparison with the literature values quoted above, we reduced our line list for \lisix\ from 9 to 3 components. We found only very minor wavelength differences of less than 1~m\AA, and an almost perfect agreement of the $\log gf$ values for all three \lisix\ components. For \liseven, there are major differences in the $\log gf$ values of the four closely spaced red HFS components. However, they are hardly relevant since the sum of their $gf$ values is very nearly the same. We also compared our lithium line list with the data given by Morton (2003), his Table 4, and found very close agreement with the Kurucz (2006) line list regrouped to six and four components for \liseven\ and \lisix, respectively.

We have double checked that the broadening parameters we used for the Li\,{\sc i} doublet are fully consistent with the ABO theory (e.g., Barklem et al. 1998) for the van der Waals broadening and with the radiative broadening given by Kurucz (2006). Stark broadening is negligible in the temperature range considered in our investigation. We present an extended version of the Mott et al. (2017) Li table in the Appendix as Table~\ref{linelist_Li} that also includes the broadening constants. 

The \xiB\ spectrum shows molecular contributions from various species. While CN lines are numerously included in the Mel\'endez et al. (2012) line list, also three C$_2$ blends, other molecular species are not. We are aware that several alternative line lists exist for many molecular species, for example, by Brooke et al. (2014) for CN. Exploring these would be a major undertaking beyond the scope of the present study. We therefore stick to the basic list of CN and C$_2$ lines provided by Mel\'endez et al. (2012), assuming that this list would properly represent the most important features due to CN and C$_2$. Table~\ref{linelist_melendez} presents the detailed list of atomic plus CN and  C$_2$ lines used for the present study, including information about the broadening constants and relevant references.

A comparison with the sunspot umbral spectrum atlas of Wallace et al. (1999) for the Li range 6707.0--6708.5\,\AA\ indicates six TiO $\gamma$-band 1-0 absorptions and one line from the CaH A-band. For the TiO $\gamma$-bands ($A^3\Phi-X^3\Delta$) wavelengths and intensities, we refer to Ram et al. (1999). With P, Q, and R being the branches of vibrational sub-bands, all line blends for the Li region are identified belonging to these branches. The individual transitions identified by Wallace et al. (1999) are

6707.2 TiO $\gamma$ 1-0 R$_2$58;
6707.3 CaH A 0-0 R$_2$48.5;

6707.4 TiO $\gamma$ 1-0 R$_3$75;
6707.4 TiO $\gamma$ 1-0 P$_2$27;

6707.8 TiO $\gamma$ 1-0 Q$_2$40;
6708.4 TiO $\gamma$ 1-0 R$_2$59; and

6708.4 TiO $\gamma$ 1-0 Q$_3$58.

Inspired by the TiO identifications, we extracted all molecular (and atomic) lines from the VALD3 line list (Ryabchikova et al. 2015) using the default configuration, which amounts to literally thousands of TiO lines, and tried to first fit the Li-spectrum of 61\,Cyg\,A. The plain VALD3 line list has $\log gf$ values of some prominent TiO lines that are partly too large, partly too small plus two prominent iron lines with grossly wrong $\log gf$ values which, together, resulted in unreasonable fits of the Li region of \xiA. Alternatively, B.~Plez privately communicated his TiO line list which is the updated version from January 2012 of the list based on Plez (1998). In McKemmish et al. (2019) this Plez-2012 line list is compared to the newer TOTO line list. At least in the wavelength range of our interest here, there seems
to be a good general agreement. We combine this line list with the Mel\'endez et al. (2012) and Lawler et al. (2014) lists (Table~\ref{linelist_melendez}) and employ it for the fit instead of the VALD3 list. It covers five titanium isotopes; $^{46}$TiO with 4542 lines, $^{47}$TiO with 4721 lines, $^{48}$TiO with 4803 lines, $^{49}$TiO with 4780 lines, and $^{50}$TiO with 4695 lines. Adding all these TiO lines still allowed only for a moderate fit for 61\,Cyg\,A and did not reveal a significant influence on the lithium fits for the initial effective temperature of \xiB\ of 4770\,K from Luck (2017). No impact whatsoever is recognizable for the fits for the even warmer \xiA . A few synthetic test runs with various effective temperatures quickly demonstrated the very steep dependency of the TiO line strength on $T_{\rm eff}$. If \xiB\ were cooler by 250\,K, the overall TiO line strength would already increase by a factor three. At $T_{\rm eff}$=4000\,K and $\log g$=5.0, both appropriate for a sunspot umbra according to Wallace et al. (1999), the maximum TiO line strength reached 30\%\ of the continuum and matches the solar umbral TiO lines around Li\,{\sc i} 6707\,\AA\ very well. Only the one TiO line at 6707.45\,\AA\ appears then too deep in the synthetic spectrum by 4\%. However, it does not blend with the Li lines but is included in the fitting range and thus contributes to the $\chi^2$.

\subsection{TiO band head effective temperature of $\xi$~Boo~B and 61~Cyg~A}

At this point we suspected that \xiB's effective temperature from Luck (2017) is too high. A comparison with a TiO spectrum of 61\,Cyg\,A (K5V) confirms this. The TiO band head absorption at 7055\,\AA\ is markedly weaker in \xiB\ than in 61\,Cyg\,A. The latter is one of the {\it Gaia} benchmark stars for which a fundamental effective temperature of 4374$\pm$22\,K and logarithmic gravity of 4.63$\pm$0.04 was derived from direct measurements of both angular diameter and bolometric flux (Heiter et al. 2015). Note though that the spectroscopic atomic-line determinations of $T_{\rm eff}$ arrived all at warmer temperatures; 4800\,K (Heiter \& Luck 2003), 4640\,K (Luck \& Heiter 2005), 4525\,K (Affer et al. 2005), and 4545\,K (Boro-Saikia et al. 2016) while an earlier determination from TiO bands suggested 4325\,K (O'Neal et al. 1998). Whether cool spots can explain such temperature differences is addressed in the Appendix.

The synthetic spectra predict a line depth $d$ of the TiO band head of $d \approx 0.45$ at \Teff=4000\,K (at solar metallicity for an average $\log g$ between 4.5 and 5.0). The observed line depth of the TiO band head is $\approx$0.11 for \xiB\ and $\approx$0.30 for 61\,Cyg\,A. By virtue of Eq.~(\ref{eq03}) in the Appendix, this translates to $\delta T = (T-4000)/1000 \approx 0.297$ and $\approx$0.101, respectively. Summarizing, we arrive at
\begin{eqnarray}
  T_{\rm eff} &\approx& 4300\,\mathrm{K\;for\;}\xi \mathrm{\;Boo\ B}\, ,
  \nonumber \\
 T_{\rm eff} &\approx& 4100\,\mathrm{K\;for\;61\;Cyg\ A}\, .
\label{eq04}
\end{eqnarray}
This is to be compared to the nominal effective temperature of these stars from the literature:
\begin{eqnarray}
  T_{\rm eff} &\approx& 4767\,\mathrm{K\;for\;}\xi \mathrm{\;Boo\ B}\, ,
  \nonumber \\
 T_{\rm eff} &\approx& 4374\,\mathrm{K\;for\;61\;Cyg\ A}\, .
\label{eq05}
\end{eqnarray}
In both cases, the \Teff\ values based on the TiO band head are significantly cooler than any of the literature  temperatures. O'Neal et al. (1998) had already shown the $T_{\rm eff}$ dependency of the band head strengths based on inactive dwarf and giants star comparison spectra. They found that TiO is most sensitive in the 3500--4000\,K range. As a compromise, we will use $T_{\rm eff}$ = 4570\,K for \xiB.

\begin{figure}
{\bf a.}\\
\mbox{\includegraphics[angle=0,width=86mm,trim=20 20 10 10]{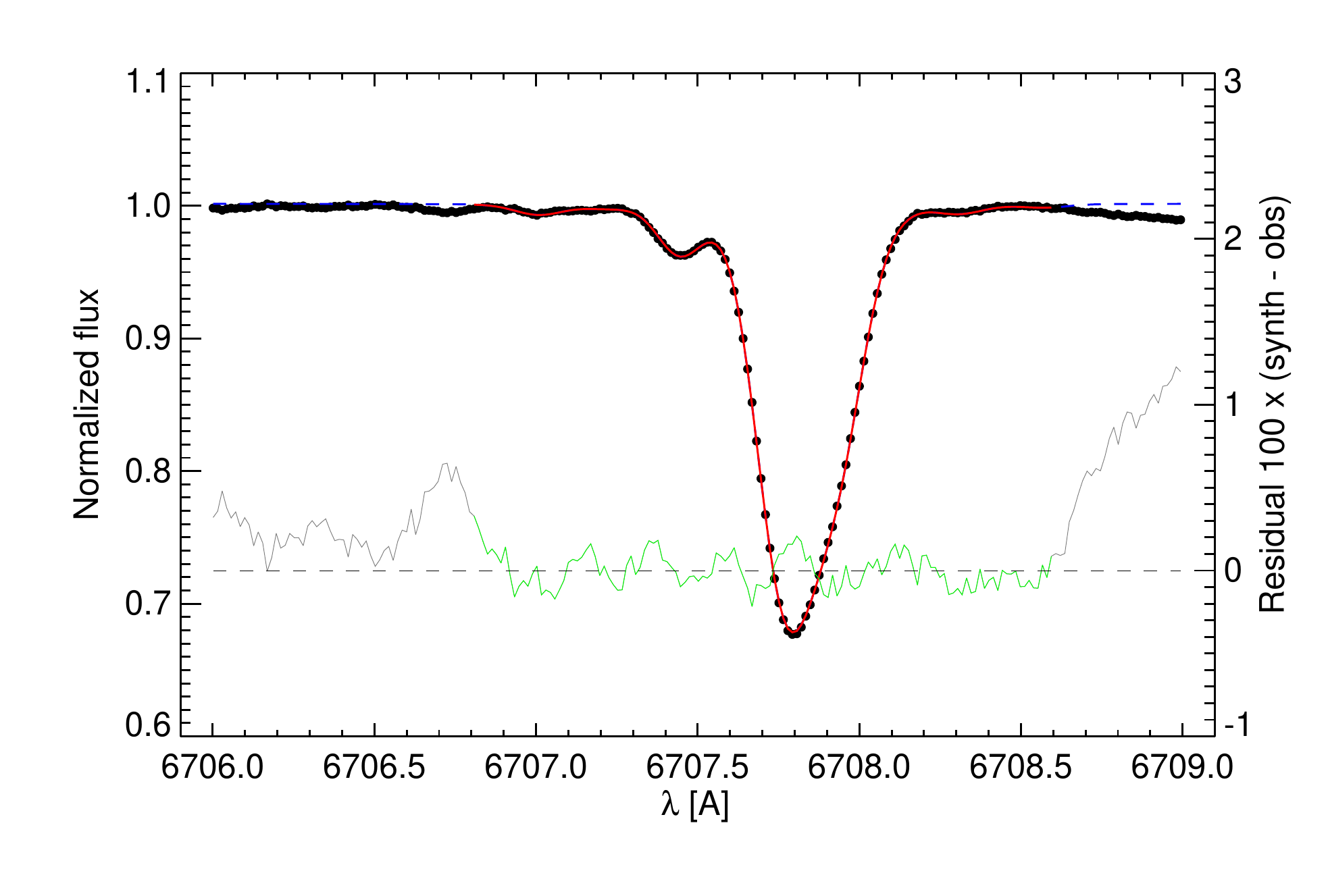}}
{\bf b.}\\
\mbox{\includegraphics[angle=0,width=86mm,trim=20 20 10 10]{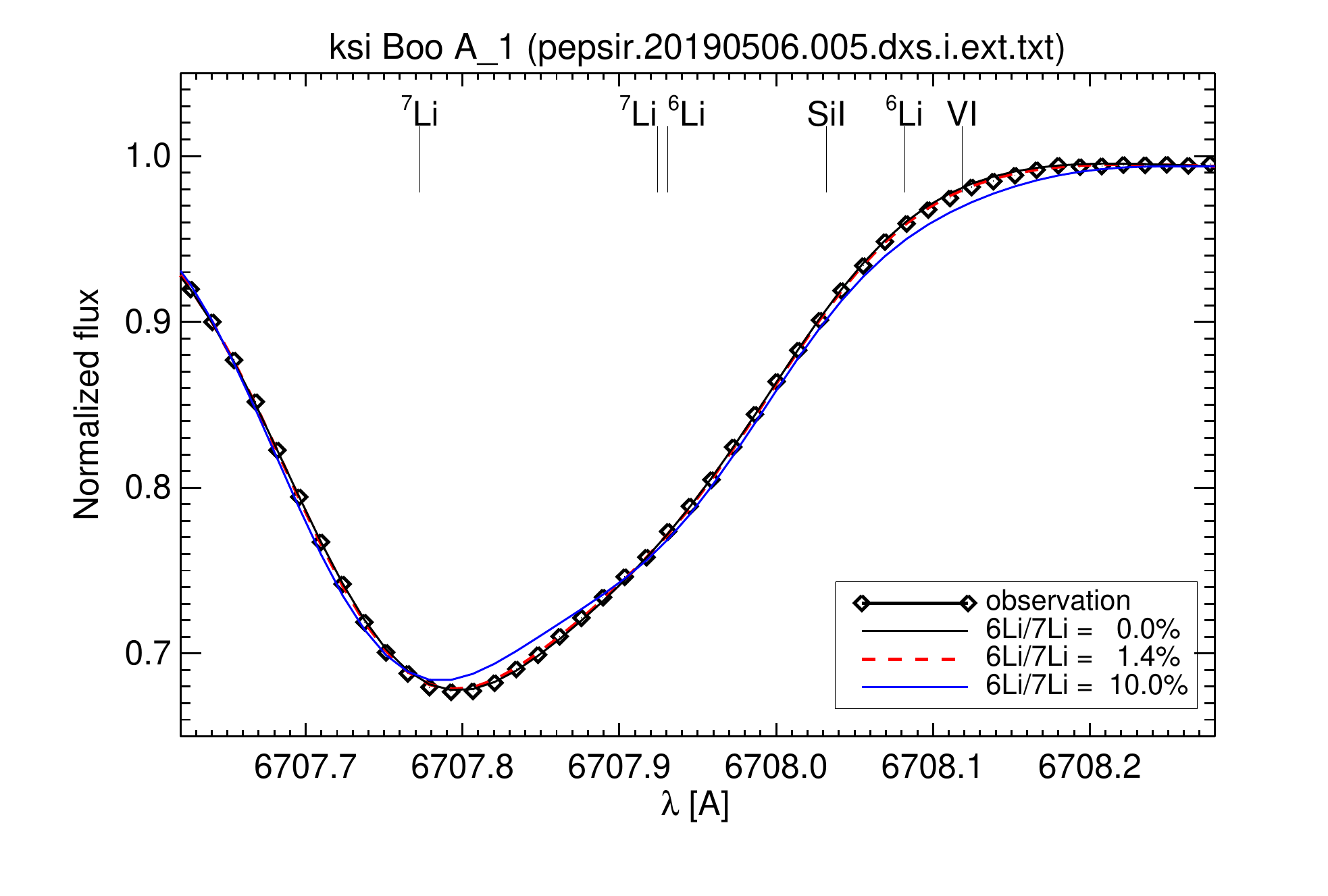}}
\caption{PEPSI spectrum of \xiA\ on JD\,2,458,609 (dots; left axes). {\it Panel a.} Best 1D-LTE fit (red line) for the full Li wavelength range. The fit residuals (line below) are expanded by a factor of 12.5 for better visibility (right axis). {\it Panel b.} Close up to the Li wavelength range with the best fit (dashed red line) and a comparison of two isotope ratios of 0\% (black) and 10\% (blue). }\label{F2}
\end{figure}

\begin{figure}
{\bf a.}\\
\mbox{\includegraphics[angle=0,width=86mm,trim=20 20 10 10]{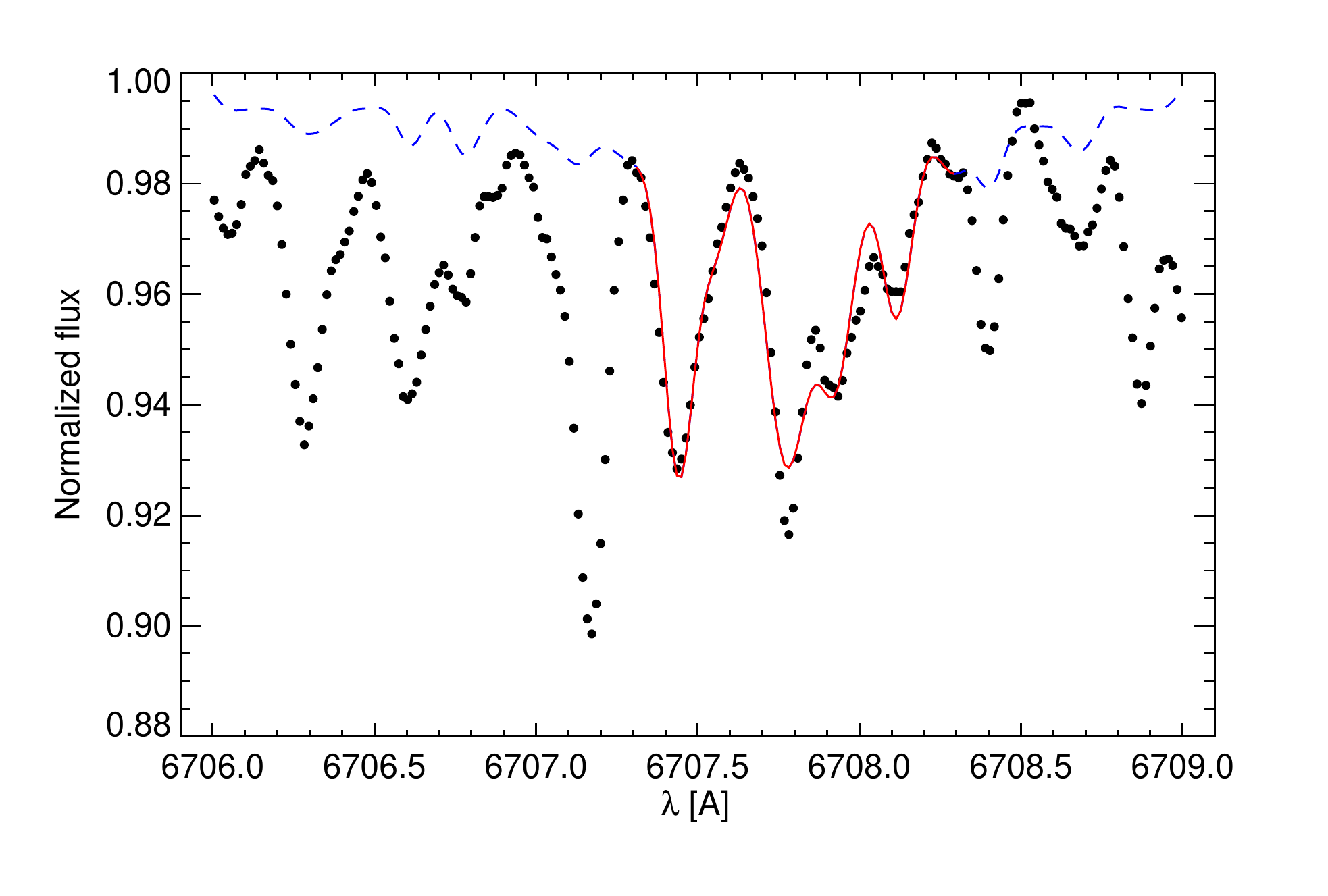}}
{\bf b.}\\
\mbox{\includegraphics[angle=0,width=86mm,trim=20 20 10 10]{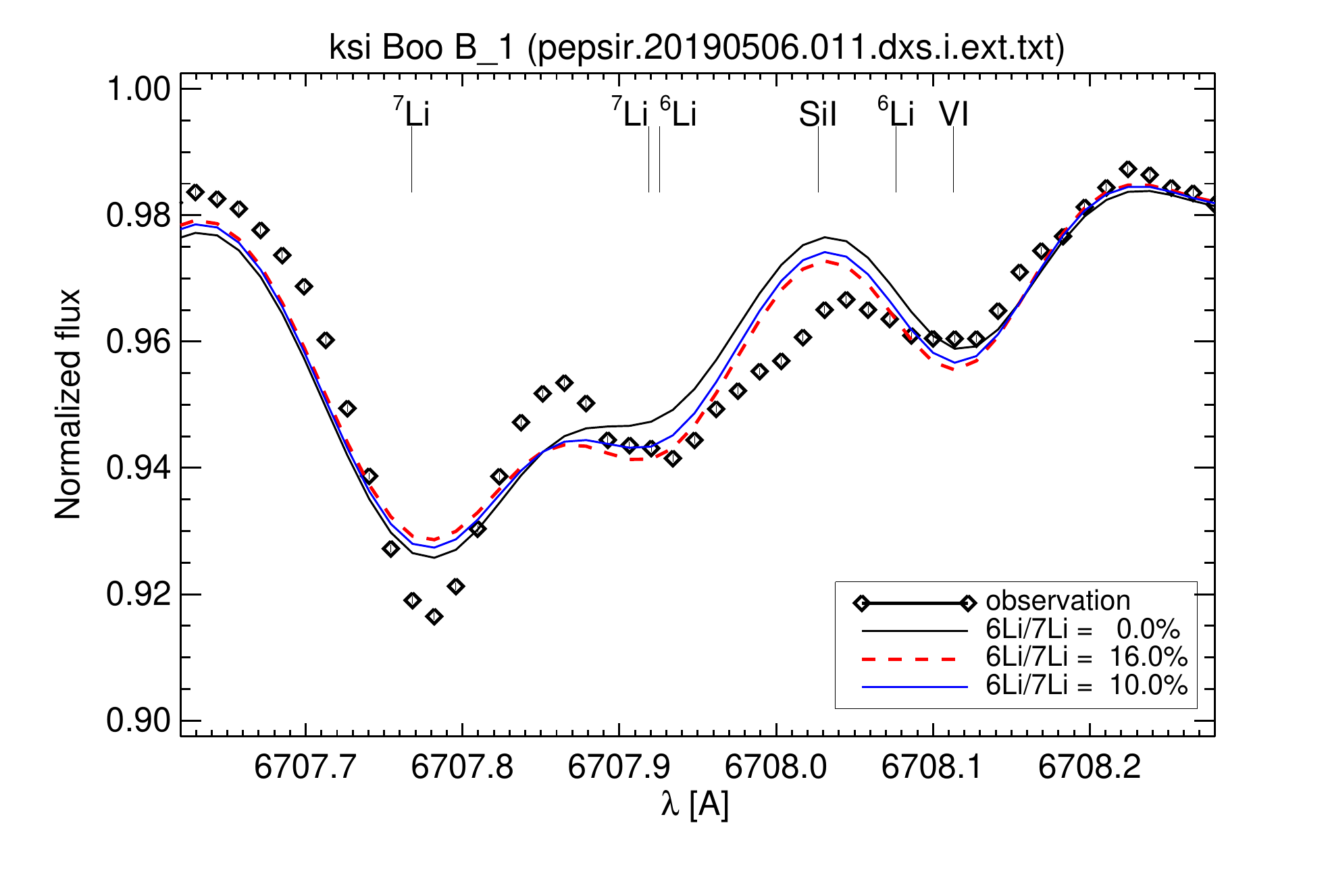}}
\caption{PEPSI spectrum of \xiB\ on JD 2,458,609 (dots). Note the reduced fitting range 6707.3--6708.3\,\AA\ (full red line). Otherwise as in Fig.~\ref{F2}.}\label{F3}
\end{figure}

\subsection{Fitting procedure}

The lithium abundance A(Li) and the $^6$Li/$^7$Li isotopic ratio are obtained by fitting the respective PEPSI spectrum with synthetic spectra obtained by interpolation from the pre-computed grid of synthetic line profiles. We employ the least-squares fitting algorithm MPFIT (Markwardt 2009; described in more detail in Steffen et al. 2015) included in an IDL program called TurboMPfit. TurboMPfit was designed specifically for the present purpose, that is providing the multi-dimensional library of synthetic spectra computed with Turbospectrum as input for MPFIT together with a list of fitting parameters that are to be optimized to find the minimum $\chi^2$.

\begin{figure}
{\bf a.}\\
\mbox{\includegraphics[clip,angle=0,width=86mm,trim=0 0 40 40]{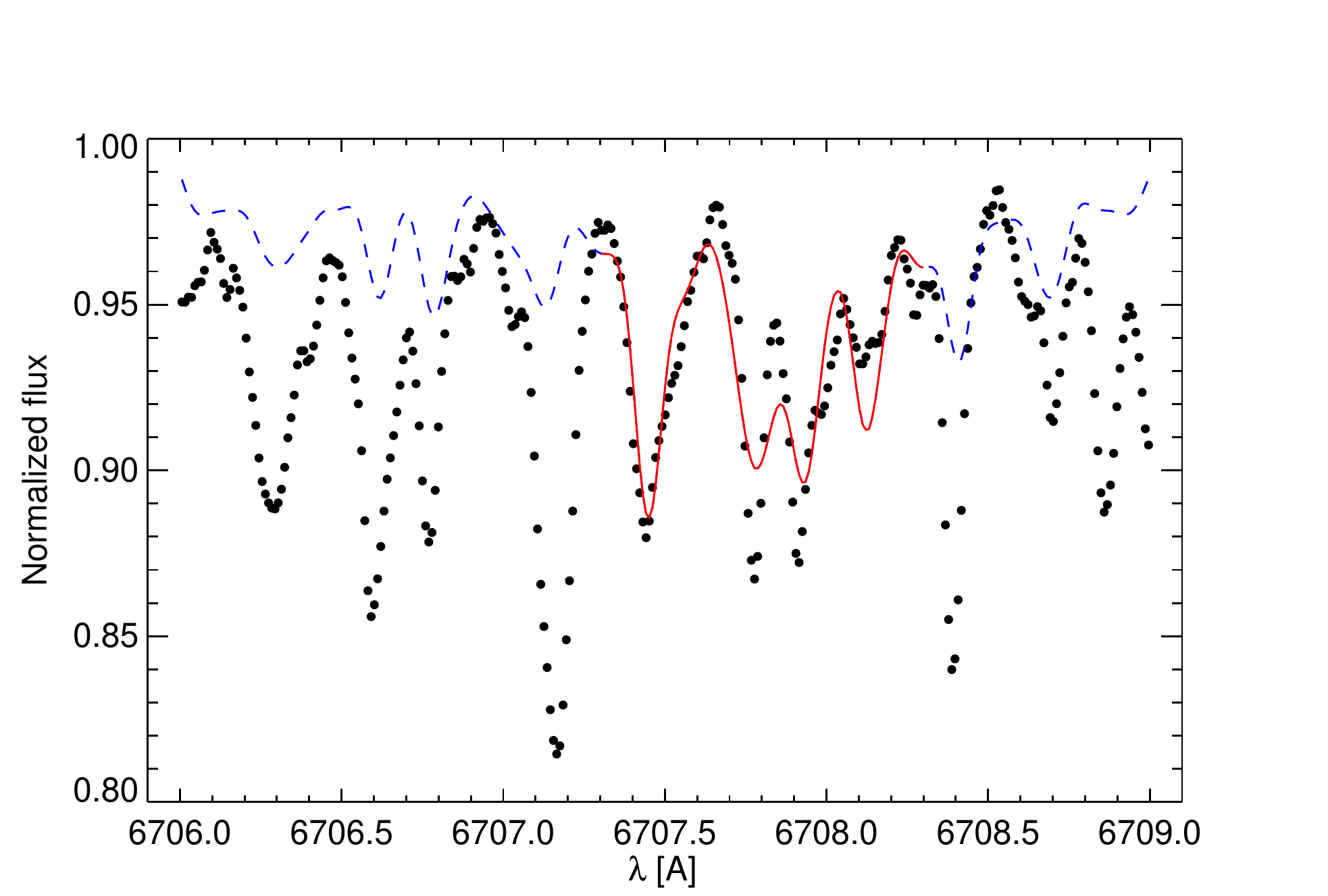}}
{\bf b.}\\
\mbox{\includegraphics[clip,angle=0,width=86mm,trim=0 0 40 40]{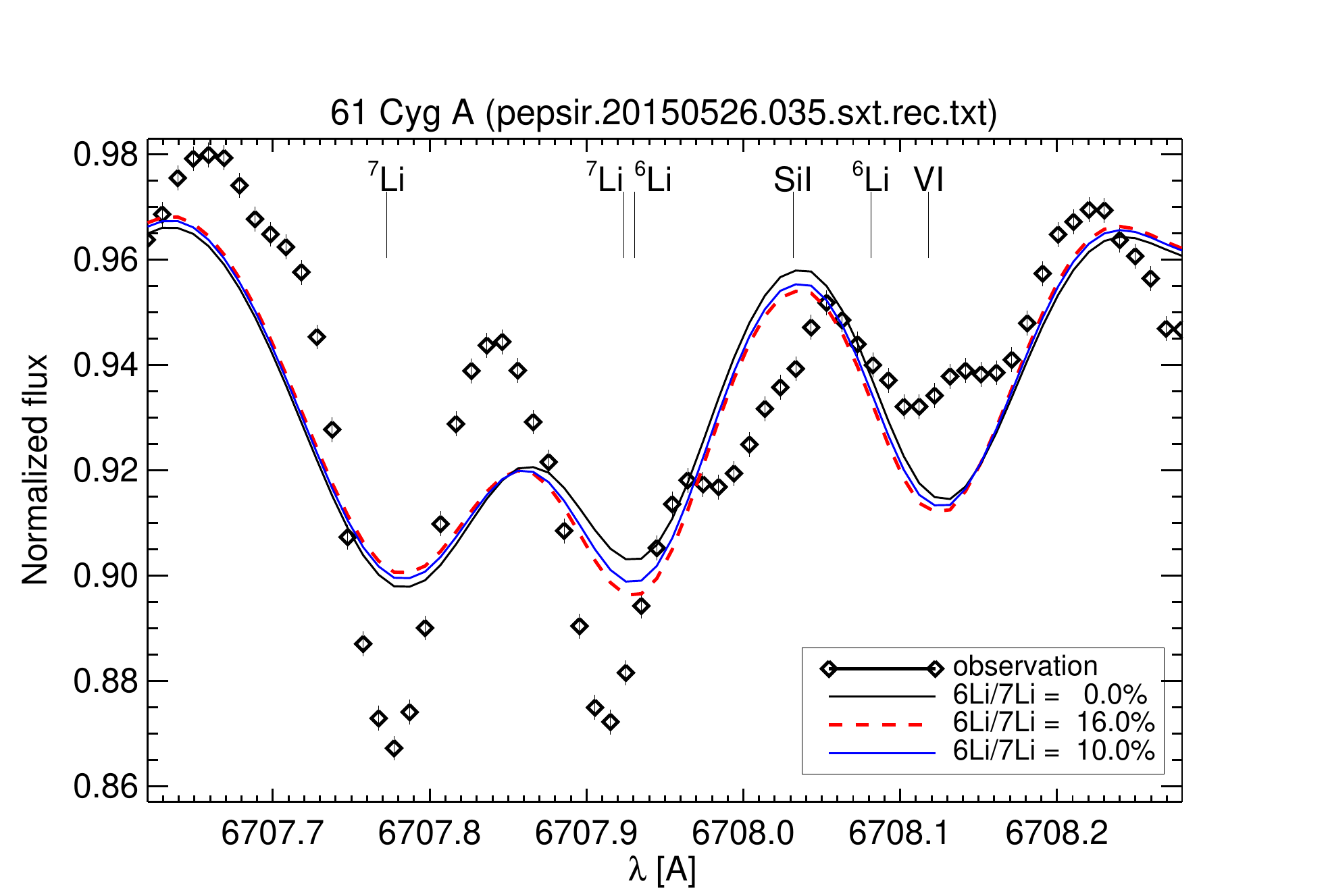}}
\caption{PEPSI spectrum of the benchmark star 61\,Cyg\,A (K5V). The fits assumed a stellar $T_{\rm eff}$ of 4374\,K. Otherwise as in Fig.~\ref{F2}.}\label{F4}
\end{figure}

The free parameters are A(Li), $^6$Li/$^7$Li, [M/H], a global wavelength adjustment, and a global Gaussian line broadening (FWHM), which are applied in velocity space to the synthetic interpolated line profiles to match the observational data as closely as possible. FWHM represents the full width half maximum of the applied Gaussian kernel and represents the combined instrumental plus macroturbulence broadening. The continuum normalization of each spectrum is iteratively optimized initially but kept fixed at the best value in the final fits. The scaled continuum shifts\footnote{Normalized fluxes are simply scaled by 1+shift.} with respect to the original data were $-0.002$ for \xiA\ and $-0.020$ for \xiB\ (and $-0.035$ for 61\,Cyg\,A). This is in particular needed for the cool targets because the many molecular lines create a suppressed quasi continuum that the data reduction software can not handle properly. We also fit the Li doublet for all our targets using two wavelength windows; a larger range of 6706.8--6708.6\,\AA\ for \xiA\ and a narrower range of 6707.3--6708.3\,\AA\ for \xiB\ and 61\,Cyg\,A. Each range includes the respective blending lines indicated in Fig.~\ref{F1}. The larger and narrower fit range is indicated in the top panels in Fig.~\ref{F2} and Fig.~\ref{F3}-\ref{F4}, respectively, by the thick red line (the spectrum outside the fit range is shown as a blue dashed line). The bottom panels of Figs.~\ref{F2}- \ref{F4} cover only the part of the fit range centered on the Li doublet. Our $\chi^2$ is computed from a 130-pixel range for \xiA\ and 72 pixels for \xiB.

\subsection{Results}

The most consistent spectrum fits for \xiA\ were achieved with stellar parameters of $T_{\rm eff}$=5480\,K, $\log g$=4.53, [M/H]=$-0.13$, $v_{\rm micro}$=1.40, and $v\sin i$=3.0\,\kms. The best fits for \xiB\ were achieved with stellar parameters of $T_{\rm eff}$=4570\,K, $\log g$=5.00, [M/H]=$+0.13$, $v_{\rm micro}$=0.15, and $v\sin i$=1.50\,\kms. In the appendix, we apply also a spot model with $T_{\rm spot}$=3800\,K and a spot-area filling factor of $a=0.3$. Note that $T_{\rm eff}$, $\log g$, and $v_{\rm micro}$ were never solved for in our analysis but assumed fixed in the input. Combined with the observed rotation periods for both stars, above $v\sin i$ values imply minimum stellar radii of 0.38\,R$_\odot$ and 0.35\,R$_\odot$ for \xiA\ and \xiB, respectively. These numbers are much smaller than the nominal radii expected from the respective spectral classifications of G8V and K5V, and already hint towards a high inclination of the rotational axes with respect to the sky. The inclination of the rotation axis of \xiA\ was indeed determined from Zeeman-Doppler imaging to $i=28\pm5$\degr\ (Petit et al. 2005, Morgenthaler et al. 2012). This converts the minimum radius for \xiA\ to a radius of 0.82\,R$_\odot$, in very good agreement with its G8V classification. No observed inclination is available for \xiB\ but, if we assume a radius of 0.66\,R$_\odot$ for a K5V star as determined for 61\,Cyg~A (K5V) by interferometry (Kervella et al. 2008), the expected inclination for \xiB\ would be around $\approx$32\degr, very similar to the observed inclination of \xiA. It is now tempting to assume that the rotational axes of both components had been co-aligned over their evolutionary history. Note that there remains an inclination of $\approx$20\degr\ with respect to AB's joint orbital plane suggested by the orbital elements from Wielen (1962). With the mass-radius relation of Demory et al. (2009), above radii yield the most-likely masses of 0.85\,M$_\odot$ and 0.68\,M$_\odot$ for \xiA\ and \xiB\, respectively. For such low masses the standard model (Iben 1967) predicts that $^6$Li is completely destroyed early in the pre-main-sequence phase, within the first two million years.

Figures~\ref{F2} and \ref{F3} show the final Li\,{\sc i} 6708-\AA\ line fits for one example spectrum for both $\xi$\,Boo stars without spot models. The results from all individual spectra are summarized in Table~\ref{T2} with the respective numerical values for the 1D-LTE case. Grand average lithium abundances of A(Li)=2.400$\pm$0.031 and 0.37$\pm$0.09 for \xiA\ and \xiB, respectively, were derived. A 3D-NLTE correction is only available for the temperatures of \xiA\ and is +0.05\,dex (Mott et al. 2020). No such correction is available for the effective temperature of \xiB . However, a simple extrapolation of the results in Mott et al. (2020) suggests a correction in the range of approximately +0.20, which would be compatible with Lind et al. (2009). Note that the error bars in Table~\ref{T2} are the formal 1$\sigma$ fitting errors on the free parameters, and do not indicate the final uncertainty of the measurement. The external errors given above for both A(Li) values were estimated from the contributions of assumed uncertainties of the stellar parameters, most notably of the effective temperature of at least $\pm$30\,K, and from our previous comparison of different line lists (Mott et al. 2017). In this way, the best estimate for the external error of A(Li) for \xiA\ and \xiB\ is 0.03\,dex and 0.09\,dex ($\approx$7\%\ and $\approx$23\% of their respective absolute abundances), respectively, roughly 30 times the internal fitting error. The above values of A(Li) make the Li abundance for the cool B-component three times less than the Sun's, while for the A-component it is 23 times higher than the abundance of the Sun (for comparison A(Li)$_{\rm Sun}$ = 1.09$\pm$0.04~dex from 3D~NLTE based on PEPSI spectra of the Sun-as-a-star; Strassmeier et al. 2018a).

Global metallicities [M/H] appear also significantly different for the two binary components. Our fits of the iron blend $\lambda$6707.426\,\AA\ suggest an average [M/H] = --0.13$\pm$0.01 for \xiA\ while +0.13$\pm$0.02 for \xiB. Like in the case of the Li abundance the internal errors are so small, mostly not larger than $\pm$0.002 for [M/H], that the actual external error is again dominated by the errors from $T_{\rm eff}$, $\log g$, and the line list. The super-solar metallicity of \xiB\ is disturbing but consistently reconstructed from the Li fits. However, as long as we face massive line list uncertainties, we consider the super-solar metallicity of \xiB\ as inconclusive.

Figures~\ref{F2}b and \ref{F3}b each show synthetic spectra with a fixed isotope ratio of 0\% and 10\% for comparison, along with the best-fit value as indicated in the plot. The deviations of the 10\% line from the data are obvious, while no difference can be seen by eye between the best fit and the 0\% assumption in case of \xiA. Note that the specific fit parameter $^6$Li/$^7$Li was limited to values no larger than 16\% for computational reasons, which was reached by the ``best fit'' in case of \xiB. However, this ratio is artificial because of the overall misfit due to yet unaccounted line contributions. Therefore, no statement regarding \xiB's $^6$Li can be made.

Finally, Fig.~\ref{F4} is a comparison with the Li region of the K5V benchmark star 61\,Cyg\,A. The spectrum was taken from Strassmeier et al. (2018b) and has $R\approx$250\,000 and S/N of 425 per pixel. Our high-resolution spectrum of \xiB\ very much resembles that of the benchmark star 61\,Cyg\,A, in agreement with its original K5V classification by Abt (1981) and its level of magnetic activity (Boro-Saikia et al. 2016). The same line list as for \xiB\ was employed. Adopted astrophysical parameters $T_{\rm eff}$, $\log g$, [Fe/H], $v_{\rm micro}$, and $v\sin i$ were 4374, 4.63, +0.15, 0.84, and 0.00 in the usual units, respectively. These parameters yielded a best fit for 61\,Cyg\,A with A(Li) = 0.146$\pm$0.004 (internal 1-$\sigma$ error). We also adopted a spot model for its Li fit with $T_{\rm spot}$=3500\,K and $a=0.5$. The best-fit Li abundance is then A(Li)=0.53$\pm$0.05 in 1D-LTE at a $\chi^2$ of 4180 for a 102-pixel range. Neither of the two approaches could fit the two $^7$Li lines to their full line depths, opposite to the V\,{\sc i} 6708.094\,\AA\ and $^6$Li 6708.07\,\AA\ blend. The reasons for this are likely inaccurate line parameters in the molecular and atomic line lists combined with an inhomogeneous surface temperature distribution, like for \xiB. We note that in the same fit the line depth of the blend Fe\,{\sc i} 6707.43\,\AA\ plus TiO 6707.45\,\AA\ was reproduced correctly (but probably at the expense of an overestimated iron abundance and macroturbulence). Significant errors in the broadening constants of the Li lines can be ruled out as an explanation for the fact that the synthetic Li lines appear to be too broad compared to the observations in the fits for \xiB\ and 61\,Cyg\,A. Rather, this is a result of the $\chi^2$ minimization procedure defining the best fit. Presumably, the line list is incomplete or incorrect on the blue side of the Li doublet, such that the best fit is achieved with an enhanced global line broadening and a somewhat increased metallicity. Nevertheless, the two fits suggest that 61\,Cyg\,A has a small amount of lithium left on its surface, comparable to \xiB.

\begin{figure}
\includegraphics[angle=0,width=86mm,clip]{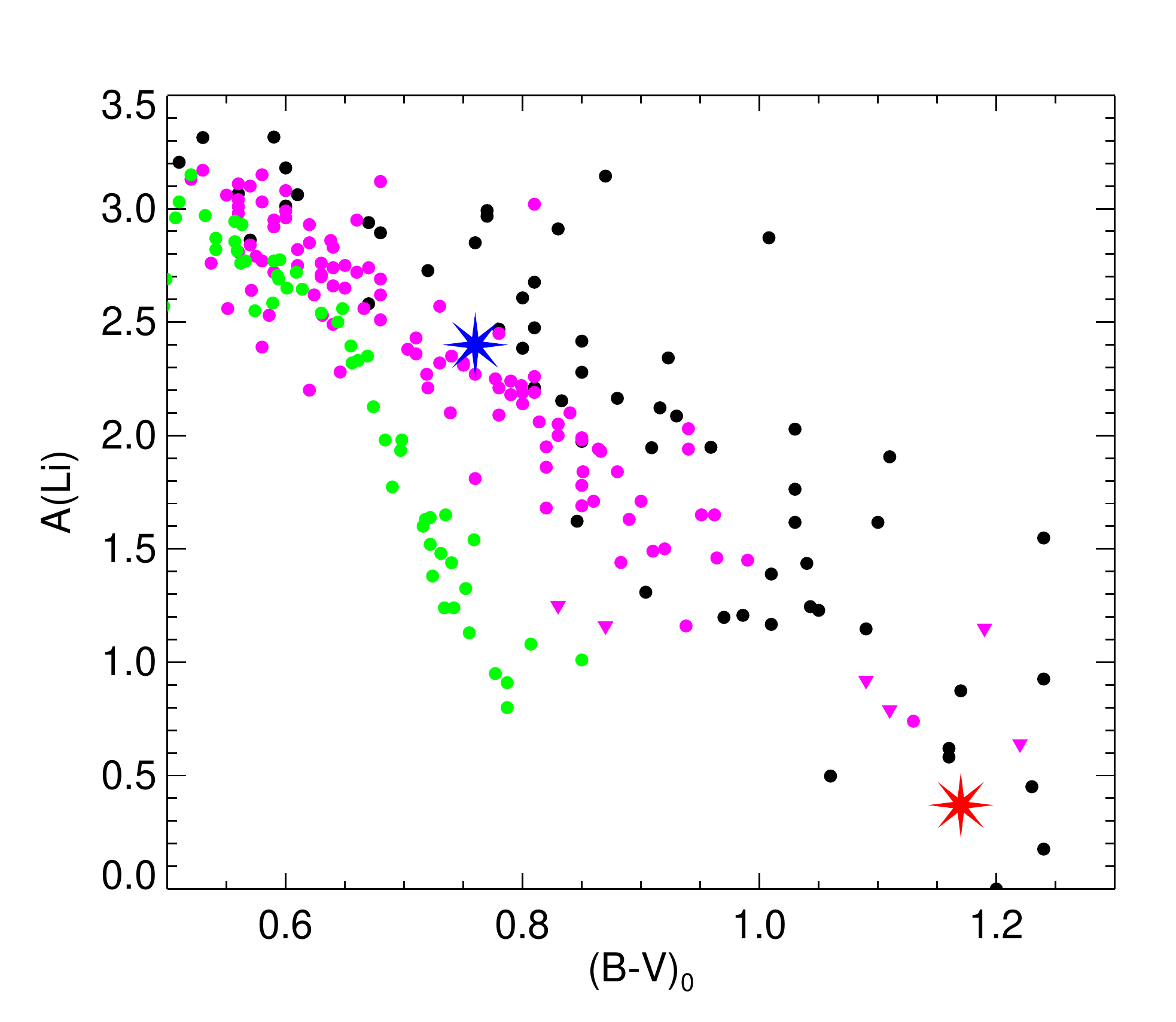}
\caption{LTE lithium abundance, A(Li), versus color $(B-V)_0$ for the binary components $\xi$~Boo A and B (blue and red star symbols, respectively), superimposed on the A(Li) vs. $(B-V)_0$ distribution of likely single members of four open star clusters: Pleiades ($\approx$125\,Myr, black dots, A(Li) from Bouvier at al. 2018; M35 ($\approx$200\,Myr, magenta dots and triangles indicating upper limits, taken from Barrado~y~Navascues et al. 2001 and Anthony-Twarog et al. 2018); and Hyades and Praesepe combined ($\approx$650\,Myr, green dots, from Cummings et al. 2017). The figure demonstrates that the lithium abundance of the $\xi$\,Boo binary components is fully consistent with the trend of similar-aged M35. }\label{F5}
\end{figure}

\subsection{Impact of spots}

Vogt (1981) had demonstrated that the appearance of a large spot on the disk of the active star II~Peg (K2\,IV) caused a steep flux rise to the red and the appearance of molecular absorption features of VO and the $\gamma$ system of TiO. The most pronounced feature was the TiO $\gamma$ bandhead at 7055\,\AA. The differential band head was later used as a star spot temperature diagnostic (O'Neal et al. 1998, 2004) and applied to numerous stellar targets.

Compared to the quiet Sun, the lithium equivalent width was shown to be dramatically weakened in bright plage regions and strengthened in dark sunspot umbrae (Giampapa 1984). While equivalent widths and line-strength changes are related to the low ionization potential of lithium and certainly have an additional non-radiative heating contribution from magnetic features, the Li abundance itself in these features was thus never conclusively determined or found to be non-uniform. Hultqvist (1974) theoretically explored spatial inhomogeneities of the solar lithium abundance that could result from spallation reactions during solar flares. Evidence for it was extensively searched for in more active stars than the Sun but never conclusively found (e.g., Randich et al. 1994). The spot activity in concert with the large S/N of our \xiA\ spectra at least enable a closer look towards this issue.

We claim that if systematic deviations of the individual abundances in our time series of eight spectra follow the expected rotational modulation, then these deviations could be due to spot enhancement. No systematic deviations above 1$\sigma$ were found though (Table~\ref{T2}), neither for component A nor component B, which places an estimated upper limit for the apparent abundance enhancement to $<0.004$\,dex.

\section{Summary and conclusions}\label{S5}

In this paper we present new spectra for both components of the visual binary $\xi$~Boo and perform a detailed analysis of their lithium abundances with the Li\,{\sc i} 6708-\AA\ feature. Our new spectra are of currently highest possible quality with a two-pixel spectral resolution of 130\,000 and S/N per pixel of up to 3200 for \xiA\ and 1700 for \xiB. Based on the line list of Mel\'endez et al. (2012), we added molecular TiO lines for the fit to the cooler B component. The same line list is also applied to a spectrum of the K5V benchmark star 61\,Cyg\,A. While the fit to the observation is done with 1D-LTE synthetic spectra, we add a 3-D NLTE correction for \xiA\ (none is available for the cool B component but is estimated to be of the order of +0.20\,dex). Our final result is that the Li abundance of the cool B-component is at least three times less than the Sun's, while for the A-component it is 23 times higher than the abundance of the Sun. 1D-LTE $^7$Li abundances were measured to A(Li) = 2.40$\pm$0.03\,dex for component~A and A(Li) = 0.37$\pm$0.09\,dex for component~B. No $^6$Li is detected for the A component, no statement is possible for the B component. At the same time the respective global metallicities also appear to be very different; the A-component has subsolar metallicity consistent with previous analyses, the B-component supersolar metallicity. While the Li difference is explainable by the different masses, effective temperatures and therefore mixing processes and efficiencies, the global metallicities should have been the same for two coeval binary components of sub-solar mass. Presumably, the metallicity of the B component is overestimated as a consequence of the incomplete line list employed for the spectrum synthesis. We consider it inconclusive at the moment. The gyrochronological age for both components is in agreement with each other and is $\approx$200~Myr. For 61\,Cyg\,A we obtained a small but non-zero amount of Li of A(Li)$\geq$0.10\,dex, which increases to 0.53\,dex if cool spots as large as a half a hemisphere are included. However, we emphasize that the lack of an adequate line list for stellar effective temperatures below $\approx$5000\,K makes quantitative conclusions about the Li abundance of \xiB\ and 61\,Cyg~A rather uncertain.

\xiA\ and \xiB's lithium abundances appear in agreement with the $T_{\rm eff}$ and age trend found for stars in the three open clusters Pleiades, M35, and Hyades (Anthony-Twarog et al. 2018). Both our measurements very well fit the trend for M35, a cluster of subsolar metallicity and an age of $\approx$200~Myr, both quantities comparable to $\xi$\,Boo. In Fig.~\ref{F5} we show the location of $\xi$\,Boo A and B in the A(Li) versus $(B-V)_0$
diagram in comparison with the lithium abundances of the various open cluster stars taken from the relevant literature. This plot clearly
demonstrates that the lithium abundance of the $\xi$\,Boo binary components is fully consistent with the trend of M35, and that the gyrochronology age of the binary obtained by Barnes (2007) agrees closely with the age of this cluster. The large Li abundance difference between the two stellar components would thus be expected. Hence, neither A(Li) nor the rotation rates of the $\xi$\,Boo system appear to show any evidence for binary interaction. The non-detection of $^6$Li for \xiA\ for a star at age $\approx$200~Myr simply hints toward a time scale for its depletion process that is shorter than that.

Our TiO band head simulations proved that the presence of star spots may have a significant impact on the TiO spectrum of cool stars like $\xi$~Boo~B and 61\,Cyg~A for both of which a substantial fraction of the stellar surface is covered by star spots. Such cool spots would certainly also affect the strength of the lithium line in a way that the Li abundance derived with a uniform model atmosphere of nominal effective temperature would likely be in error. It is therefore more appropriate to fit the Li region of stars like $\xi$~Boo~B or 61\,Cyg~A with a simple two-temperature spot model. Because it would not only be cool spots that contribute to a line profile (but also bright faculae or dark granulation lanes, or disk-projected prominences, a.o.), such a two-component fit may not necessarily result in a better overall fit of the observed spectrum and thus in an improved estimate of the Li abundance. Nevertheless, our Li fits for the two cool dwarfs include a pre-set spot model with reasonable spot temperatures and spot coverage. The abundance difference $\Delta$A(Li)  from a fit with and without a spot model is only small though. For \xiB\ the difference is 0.1~\,dex (higher abundance in case with a spot contribution) but, most important, it does not improve the fit quality. The same is true for 61\,Cyg\,A where $\Delta$A(Li) amounts to +0.4\,dex. At this point, we conclude that it is the line list that must be improved with many more molecules than TiO, for example, CaH, FeH, CN, a.o.. For a cool star heavily covered by magnetic-activity tracers, 3D non-LTE abundance corrections are probably a second order effect.

\subsection{Acknowledgements} We thank Ilya Ilyin for his observational and data reduction skills and Michi Weber for extracting the VALD3 line list for us. Bertrand Plez is thanked for sharing his TiO line list with us. We also thank the referee, Ulrike Heiter, for the many helpful comments that clearly improved the paper and also for the zillions of detailed hints on the various line lists employed. LBT Corporation partners are the University of Arizona on behalf of the Arizona university system; Istituto Nazionale di Astrofisica, Italy; LBT Beteiligungsgesellschaft, Germany, representing the Max-Planck Society, the Leibniz-Institute for Astrophysics Potsdam (AIP), and Heidelberg University; the Ohio State University; and the Research Corporation, on behalf of the University of Notre Dame, University of Minnesota and University of Virginia. It is a pleasure to thank the German Federal Ministry (BMBF) for the year-long support for the construction of PEPSI through their Verbundforschung grants 05AL2BA1/3 and 05A08BAC as well as the State of Brandenburg for the continuing support of AIP and PEPSI (see https://pepsi.aip.de/). This work has made use of the VALD database, operated at Uppsala University, the Institute of Astronomy RAS in Moscow, and the University of Vienna.

\appendix

\section{Line lists}

\begin{table*}[!htb]
  \caption{Twelve-component representation of the \liseven\ and \lisix\
    isotopic hyper-fine structure of the lithium resonance doublet used in
    this work. The 19 HFS components given by \citet{kurucz06} were reduced
    to 12 components by combining, for each isotope and fine structure
    component separately, hyperfine transitions of identical wavelength.}
\label{linelist_Li}
\centering
\begin{tabular}{cccccccc}
\noalign{\smallskip}\hline\noalign{\smallskip}
\multicolumn{4}{c}{\textbf{Lithium HFS}} \\ \hline
  Wavelength & Lithium & Excitation & log $gf$ & $\sigma_{\rm ABO}$ &
  $\alpha_{\rm ABO}$ & log $(\gamma_4/N_e)^{a)}$ & log $\gamma_{\rm rad}$
  \\
  $\lambda$ (\AA) & isotope & potential (eV) & (dex) & (a.u.) &
  & (cm$^3$s$^{-1}$) & (rad s$^{-1}$)\\ \hline
6707.756 & \liseven & 0.000 & $-$0.428 & 346 & 0.236 & $-$5.78 & 7.567 \\ 
6707.768 & \liseven & 0.000 & $-$0.206 & 346 & 0.236 & $-$5.78 & 7.567 \\ 
6707.907 & \liseven & 0.000 & $-$0.808 & 346 & 0.236 & $-$5.78 & 7.567 \\ 
6707.908 & \liseven & 0.000 & $-$1.507 & 346 & 0.236 & $-$5.78 & 7.567 \\ 
6707.919 & \liseven & 0.000 & $-$0.808 & 346 & 0.236 & $-$5.78 & 7.567 \\ 
6707.920 & \liseven & 0.000 & $-$0.808 & 346 & 0.236 & $-$5.78 & 7.567 \\
  \hline\noalign{\smallskip}
6707.920 & \lisix   & 0.000 & $-$0.479 & 346 & 0.236 & $-$5.78 & 7.567 \\ 
6707.923 & \lisix   & 0.000 & $-$0.178 & 346 & 0.236 & $-$5.78 & 7.567 \\ 
6708.069 & \lisix   & 0.000 & $-$0.831 & 346 & 0.236 & $-$5.78 & 7.567 \\ 
6708.070 & \lisix   & 0.000 & $-$1.734 & 346 & 0.236 & $-$5.78 & 7.567 \\ 
6708.074 & \lisix   & 0.000 & $-$0.734 & 346 & 0.236 & $-$5.78 & 7.567 \\ 
6708.075 & \lisix   & 0.000 & $-$0.831 &346 & 0.236  & $-$5.78 & 7.567 \\
  \noalign{\smallskip}\hline\noalign{\smallskip}
\end{tabular}
\begin{tablenotes}
  Notes: $^{a)}$ at $T=10\,000$\,K;
  van der Waals broadening parameters $\sigma_{\rm ABO}$ and  $\alpha_{\rm ABO}$
  from \citet{barklem00}; all other data from references specified
  in \citet{kurucz06}.
\end{tablenotes}
\end{table*}

\begin{table*}[!htb]
  \caption{List of atomic and molecular lines in the Li\, 6707\AA\ region.
    Wavelength, excitation potential, and $\log gf$ values adopted from \citet{melendez12}, except for the
    \ion{V}{i} line. For atomic lines (except  \ion{V}{i}), the van der Waals broadening parameter $\gamma_6$
    is computed by the Uns\"old approximation with line-specific enhancement factors; Stark broadening is ignored.}
\label{linelist_melendez}
\centering
\begin{tabular}{cccclccr}
\noalign{\smallskip}\hline\noalign{\smallskip}
  Wavelength & Chemical & Excitation & $\log gf$ &
  References & $\log (\gamma_6/N_{\rm H})^{\,a)}$ & $f_{\rm damp}^{\,b)}$ & log $\gamma_{\rm rad}^{\,c)}$ \\
  $\lambda$ (\AA) & species & pot. (eV) & (dex) &
  for $\lambda$ and $\log gf$  & (cm$^3$ s$^{-1}$) & & (rad s$^{-1}$) \\
  \hline\noalign{\smallskip}
6707.000 & \ion{Si}{i}  & 5.954 & --2.560 & Mandell et al. (2004)$^{\,1)}$  & $-6.812$ & 1.3 & 8.146 \\ 
6707.172 & \ion{Fe}{i}  & 5.538 & --2.810 & Mandell et al. (2004)$^{\,1)}$  & $-6.919$ & 1.6 & 8.176 \\ 
6707.433 & \ion{Fe}{i}  & 4.608 & --2.250 & Mandell et al. (2004)$^{\,1)}$  & $-7.306$ & 1.5 & 8.301 \\ 
6707.473 & \ion{Sm}{ii} & 0.933 & --1.910 & Xu et al. (2003)              & $-7.575$ & 1.0 & 5.483 \\ 
6707.596 & \ion{Cr}{i} 	& 4.208 & --2.667 & Mandell et al. (2004)$^{\,1)}$  & $-6.784$ & 10.0 & 7.176 \\ 
6708.023 & \ion{Si}{i}  & 6.000 & --2.800 & Kurucz (2007, VALD)           & $-6.762$ & -- & 8.146 \\ 
6708.110 & \ion{V}{i} 	& 1.218 & --2.630 & Lawler et al. (2014)          &  331.245$^{\,d)}$ & -- & 7.602 \\ 
6708.099 & \ion{Ce}{ii} & 0.701 & --2.120 & Palmeri et al. (2000)         & $-7.576$ & 1.0 & 4.972 \\ 
6708.282 & \ion{Fe}{i}  & 4.988 & --2.700 & Mandell et al. (2004)$^{\,1)}$   & $-7.188$ & 1.3 & 8.672 \\ 
6708.347 & \ion{Fe}{i}  & 5.486 & --2.580 & Mandell et al. (2004)$^{\,1)}$   & $-6.954$ & 1.5 & 7.968 \\ 
6708.534 & \ion{Fe}{i}  & 5.558 & --2.936 & Mandell et al. (2004)$^{\,1)}$   & $-6.905$ & 1.4 & 8.301 \\ 
6708.577 & \ion{Fe}{i}  & 5.446 & --2.684 & Mandell et al. (2004)$^{\,1)}$   & $-6.979$ & 1.4 & 8.477 \\
  \noalign{\smallskip}\hline\noalign{\smallskip}
6707.205 & CN 	            & 1.970 & --1.222 & Mandell et al. (2004)$^{\,2)}$ &  --$^{\,e)}$ & & 6.176 \\ 
6707.272 & CN 		    & 2.177 & --1.416 & Mel\'endez \& Barbuy (1999) &  -- & & 6.176 \\ 
6707.282 & CN 		    & 2.055 & --1.349 & Mel\'endez \& Barbuy (1999) &  -- & & 6.176 \\ 
6707.300 & C$_{2}$ 	    & 0.933 & --1.717 & Mel\'endez \& Cohen (2007), &  -- & & 6.771 \\ 
         &           	    &       &        & Mel\'endez \& Asplund (2008) &  -- & & \\ 
6707.371 & CN 		    & 3.050 & --0.522 & Mandell et al. (2004)$^{\,3)}$ &  -- & & 6.176 \\ 
6707.460 & CN 	            & 0.788 & --3.094 & Mel\'endez \& Barbuy (1999)  &  -- & & 6.176 \\ 
6707.461 & CN               & 0.542 & --3.730 & Mel\'endez \& Barbuy (1999)  &  -- & & 6.176 \\ 
6707.470 & CN 		    & 1.880 & --1.581 & Mandell et al. (2004)$^{\,2)}$ &  -- & & 6.176 \\ 
6707.548 & CN 		    & 0.946 & --1.588 & Mel\'endez \& Barbuy (1999)  &  -- & & 6.176 \\ 
6707.595 & CN	 	    & 1.890 & --1.451 & Mandell et al. (2004)$^{\,2)}$ &  -- & & 6.176 \\ 
6707.645 & CN 		    & 0.946 & --3.330 & Mel\'endez \& Barbuy (1999)  &  -- & & 6.176 \\ 
6707.660 & C$_{2}$ 	    & 0.926 & --1.743 & Mel\'endez \& Cohen (2007),  &  -- & & 6.771 \\ 
         &           	    &       &        & Mel\'endez \& Asplund (2008) &  -- & & \\ 
6707.809 & CN 		    & 1.221 & --1.935 & Mel\'endez \& Barbuy (1999) &  -- & & 6.176 \\ 
6707.848 & CN 		    & 3.600 & --2.417 & Mandell et al. (2004)$^{\,3)}$ &  -- & & 6.176\\ 
6707.899 & CN 		    & 3.360 & --3.110 & Mandell et al. (2004)$^{\,3)}$ &  -- & & 6.176\\ 
6707.930 & CN 		    & 1.980 & --1.651 & Mandell et al. (2004)$^{\,3)}$ &  -- & & 6.176\\ 
6707.970 & C$_{2}$ 	    & 0.920 & --1.771 & Mel\'endez \& Cohen (2007), &  -- & & 6.771 \\ 
         &           	    &       &        & Mel\'endez \& Asplund (2008) &  -- & & \\ 
6707.980 & CN 		    & 2.372 & --3.527 & Mel\'endez \& Barbuy (1999) &  -- & & 6.176 \\ 
6708.026 & CN 		    & 1.980 & --2.031 & Mandell et al. (2004)$^{\,3)}$ &  -- & & 6.176 \\ 
6708.147 & CN 		    & 1.870 & --1.884 & Mandell et al. (2004)$^{\,3)}$ &  -- & & 6.176 \\ 
6708.315 & CN 		    & 2.640 & --1.719 & Mandell et al. (2004)$^{\,3)}$ &  -- & & 6.176 \\ 
6708.370 & CN 		    & 2.640 & --2.540 & Mandell et al. (2004)$^{\,3)}$ &  -- & & 6.176 \\ 
6708.420 & CN 		    & 0.768 & --3.358 & Mel\'endez \& Barbuy (1999) &  -- & & 6.176 \\ 
6708.541 & CN 		    & 2.500 & --1.876 & Mandell et al. (2004)$^{\,3)}$ &  -- & & 6.176 \\ 
  \noalign{\smallskip}\hline\noalign{\smallskip}
\end{tabular}
\begin{tablenotes}
  $^{ a)}$ at $T=10\,000$\,K;
  $^{ b)}$ van der Waals damping enhancement factor relative to VALD3 (2022) where applicable;
  $^{ c)}$ Radiative damping constants $\gamma_{\rm rad}$ are from the VALD3
  database (2021);
  $^{ d)}$ notation for $\sigma_{\rm ABO}=331$ and  $\alpha_{\rm ABO}=0.245$;
  $^{ e)}$ no van der Waals broadening is applied for molecular lines;
  $^{ 1)}$ $\lambda$ and/or $\log gf$ presumably adjusted by Mel\'endez et al. \citet{melendez12} to
  better reproduce the solar spectrum;
  $^{ 2)}$ see also Kotlar et al. (1989);
  $^{ 3)}$ see also J{\o}rgensen \& Larsson (1990).
\end{tablenotes}
\end{table*}


\section{Observing log}

Table~\ref{T1-App} is the observing log for the $\xi$\,Boo spectra in this paper.

\begin{table*}[!tbh]
\caption{Observing log.}\label{T1-App}
\center
\begin{tabular}{lllllll}
\noalign{\smallskip}\hline\noalign{\smallskip}
 BJD mid exp.& $\phi$ & $t_{\rm exp}$ & $\Delta t$ & $\Delta\lambda$ (blue; red) & \multicolumn{2}{c}{S/N p. pixel}\\
 (+2450000)  & (Eq.\,1) & (min)    & (min)      & (\AA )                   & blue & red \\
\noalign{\smallskip}\hline\noalign{\smallskip}
\xiA : & & & & & & \\
 8609.7427427 & 0.829 & 5 & 38 & 4800-5441; 6278-7419 & 2150 & 2980 \\
 8610.8282148 & 0.998 & 5 & 38 & 4800-5441; 6278-7419 & 2020 & 2910 \\
 8611.7681903 & 0.144 & 5 & 38 & 4800-5441; 6278-7419 & 1590 & 2390 \\
 8616.8333173 & 0.932 & 5 & 38 & 4800-5441; 6278-7419 & 1170 & 1770 \\
 8617.6923984 & 0.066 & 5 & 38 & 4800-5441; 6278-7419 & 2120 & 2850 \\
 8617.8769492 & 0.094 & 5 & 38 & 4800-5441; 6278-7419 & 2450 & 3380 \\
 8619.7004465 & 0.378 & 5 & 38 & 4800-5441; 6278-7419 & 1970 & 2800 \\
 8619.8758547 & 0.405 & 5 & 38 & 4800-5441; 6278-7419 & 1370 & 3360 \\
\xiB : & & & & & & \\
 8609.7860453     & 0.124 & 10 & 68 & 4800-5441; 6278-7419 & 990 & 1850 \\
 8610.8816143$^a$ & 0.215 & 10 & 68 & 4800-5441; 6278-7419 & 300 &  240 \\
 8611.8163769     & 0.294 & 10 & 68 & 4800-5441; 6278-7419 & 890 & 1720 \\
 8616.8730350     & 0.717 & 10 & 68 & 4800-5441; 6278-7419 & 870 & 1710 \\
 8617.8435912$^b$ & 0.798 & 5  & 38 & 4800-5441; 6278-7419 & 710 & 1470 \\
 8619.7390032     & 0.957 & 10 & 68 & 4800-5441; 6278-7419 & 560 & 1080 \\
 \noalign{\smallskip}\hline
\end{tabular}
\begin{tablenotes}
The first column gives the barycentric Julian date for the time of mid exposure for Stokes I. The second column is the rotational phase $\phi$ based on the respective ephemeris for \xiA\ and for \xiB. Six individual exposures with an exposure time $t_{\rm exp}$ and a total cadence of $\Delta t$ (both in minutes) make up one phase spectrum. S/N is per pixel and is an average from  the respective wavelength region $\Delta\lambda$ (dubbed blue and red). $a$. Bad weather. $b$. Exposure time was accidentally set to 4\,min in the blue, and to 5\,min in the red.
\end{tablenotes}
\end{table*}

\section{Can star spots explain enhanced TiO absorption?}

In the observed spectra of \xiB\ and 61\,Cyg\,A, the TiO lines including the TiO band head appear significantly stronger than expected from synthetic spectra computed for the nominal stellar effective temperature. Here we use synthetic spectra of the TiO 1-0 and 0-0 bands in the $\gamma$ system with band heads near 6650\,\AA\ and 7055\,\AA , respectively, to investigate whether star spots can be responsible for this mismatch.

\subsection{Spot temperature and filling factor}

A simple model assumes that we know the true effective temperature of the star, including its spots. This is in fact the case for 61~Cyg~A where \Teff\ is known from fundamental relations (\Teff = $4374\pm 22$\,K). We then have the following relation between the fundamental \Teff\ $=T_0$, the spot temperature $T_1$, the photospheric temperature $T_2$, and the area spot filling factor $a$ in relative units of the visible stellar hemisphere
\beq
a\ T_1^4 + (1-a) T_2^4 = T_0^4 \, .
\label{eq01}
\eeq
For example, assuming $T_0 = 4374$\,K like for 61\,Cyg\,A, a spot temperature $T_1 = 4000$\,K, and $a=0.3$, we find $T_2 = 4508$\,K. This value is indeed closer to the best spectroscopically determined temperatures for 61\,Cyg\,A. Note that the Zeeman-Doppler images of 61\,Cyg\,A in Boro-Saikia et al. (2016) can give no information on the temperature filling factor because the spectral lines are just barely broadened by rotation ($v\sin i\approx 1$\,\kms ) and thus do not provide spatial resolution via the Doppler effect. However, it showed magnetic fields covering large fractions of the stellar disk suggesting also a large (areal) spot filling factor.

\subsection{Continuum flux and strength of TiO features}

\begin{figure*}
\centering
\includegraphics[angle=0,width=145mm]{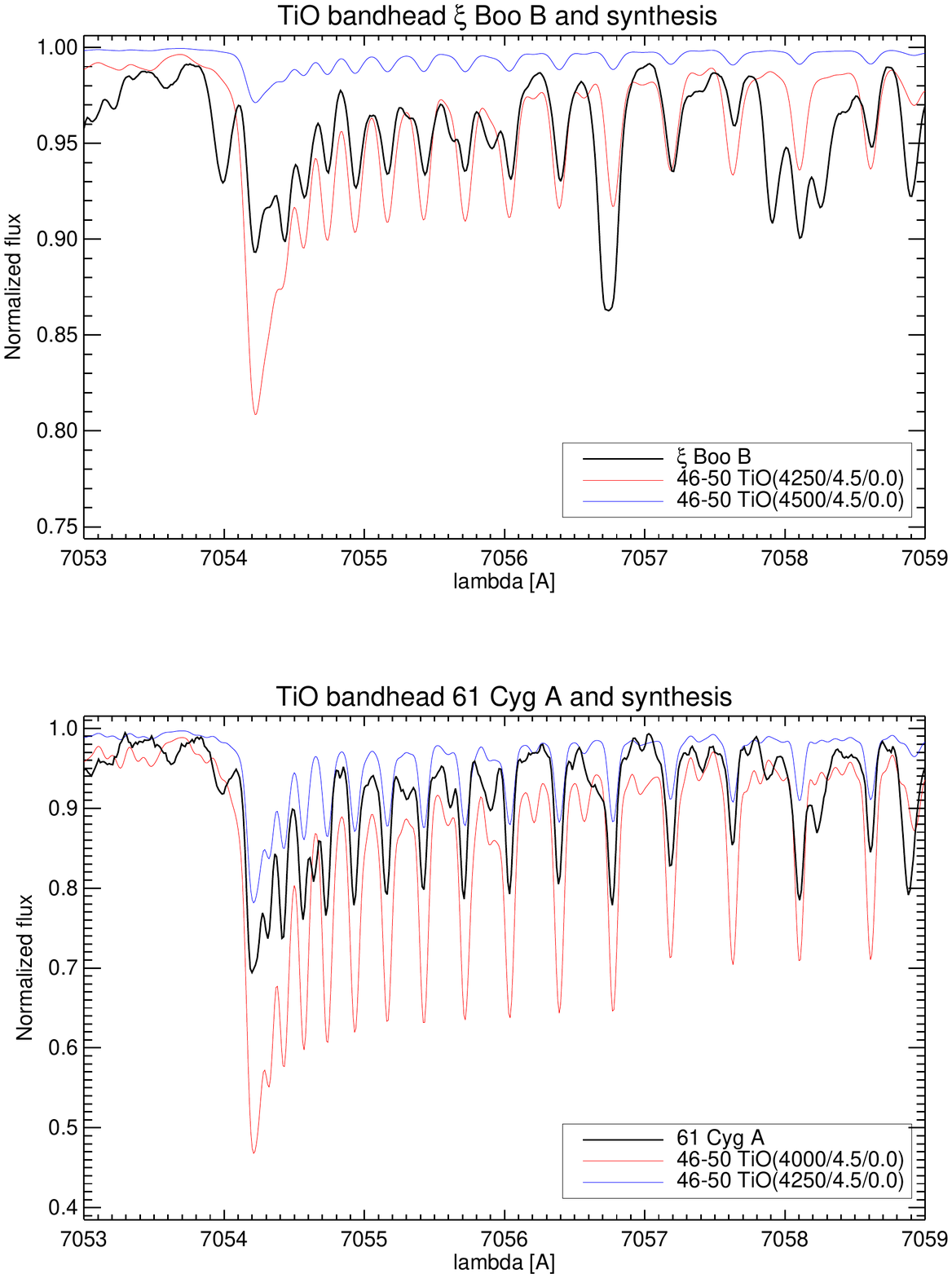}
\caption{Comparisons of observed and synthetic spectra of the TiO $\gamma$ 0-0 band
  head at $\lambda\,7055$\,\AA. For comparison with $\xi$~Boo~B (top panel, thick line)
  and 61~Cyg~A (bottom panel, thick line), the synthetic spectra have been broadened with
  a Gaussian of FWHM=4.0 and 2.0\,km/s, respectively, roughly representing
  the effect of instrumental broadening plus macro turbulence. The effective
  temperature of the superimposed synthetic spectra are chosen to bracket the
  observed strength of the TiO lines. In all cases, we adopt $\log g=4.5$
  and solar metallicity.}\label{FA1}
\end{figure*}

From synthetic TiO spectra computed with Turbospectrum (Plez 2012), we can derive (i) a relation between the monochromatic continuum flux and effective temperature and, (ii), a scaling relation between the normalized TiO line depth and \Teff. The following numerical relations are valid for $\log g=4.75$. For the continuum flux, we find
\begin{eqnarray}
\frac{F_\lambda(T)-F_\lambda(4\,\mathrm{kK})}{F_\lambda(4\,\mathrm{kK})} & \approx &
1.231\,\delta T + 0.694\,(\delta T)^2 \nonumber
\end{eqnarray}
for $\lambda$ = 6707\,\AA , and
\begin{eqnarray}
\approx\ 1.153\,\delta T + 0.615\,(\delta T)^2  \label{eq02}
\end{eqnarray}
for $\lambda$ = 7055\,\AA\ where $\delta T = (T-4000)/1000$. For the normalized depth $d$ of (any) TiO spectral feature, we find
\begin{eqnarray}
\log_{10}\left(\frac{d(T)}{d(4\,\mathrm{kK})}\right) \ \approx \hspace{20mm} \nonumber \\
-1.75\,\delta T - 1.38\,(\delta T)^2 + 0.200\,(\delta T)^3 + 0.369\,\,(\delta T)^4 \nonumber
\end{eqnarray}
averaged over the wavelength range $\lambda = 6708.4\pm 0.1$\,\AA, and
\begin{equation}
-1.57\,\delta T - 1.55\,(\delta T)^2 - 0.492\,(\delta T)^3 + 0.658\,\,(\delta T)^4  \label{eq03}
\end{equation}
averaged over the wavelength range $\lambda = 7055\pm 1$\,\AA. The latter equation can be used to derive a formal effective temperature from the strength of the TiO band head. Figure~\ref{FA1} is a direct comparison of the band head strengths for \xiB\ and 61\,Cyg\,A with synthetic spectra for various effective temperatures.

\subsection{Starspot contribution}

We investigate a simple model assuming that a fraction $a$ of the stellar surface is covered by spots with a lower temperature $T_1$, while the rest of the stellar surface radiates with effective temperature $T_2$. Using the above relations, we can then estimate the strength of the TiO absorption in the combined spectrum, $\overline{d}$. The following relation holds for the line depth of the combined spectrum
\beq
\overline{d} = \frac{a\,F_{\lambda,1}\,d_1 +  (1-a)\,F_{\lambda,2}\,d_2}
         {a\,F_{\lambda,1} +  (1-a)\,F_{\lambda,2}} =
         \frac{a\,R_{\lambda}\,d_1 +  (1-a)\,d_2}{a\,R_{\lambda} +  (1-a)}\, ,
\label{eq06}
\eeq
where $F_{\lambda,1}$ and  $F_{\lambda,2}$ denote the monochromatic continuum flux in the spots and the quiet photosphere, respectively, $R_{\lambda}=F_{\lambda,1}/F_{\lambda,2}$, and $d_1$ and $d_2$ refer to the normalized line depth in the local spectra of the two components. Note that completely dark spots ($R_{\lambda}=0$) would not have any impact on the combined spectrum, $\overline{d} = d_2$. Obviously, there is a certain optimum spot temperature where the spectral signature of the star spots is at its maximum.

Given the mean (true) effective temperature of the star, \Teff\ = $T_0$, we can now compute $\overline{d}$ for any combination of $a$ and $T_1$.  First we obtain $T_2$ from Eq.\,(\ref{eq01}). Then the continuum fluxes $F_{\lambda,1}$ and  $F_{\lambda,2}$ are obtained from Eq.\,(\ref{eq02}) and the local line depths from  Eq.\,(\ref{eq03}). Finally, these quantities are used to evaluate $\overline{d}$ from Eq.\,(\ref{eq06}).

\subsubsection{Results for 61~Cyg~A}

Adopting $T_{\rm eff}$=4374\,K, the observed strength of the TiO band head ($\overline{d}\approx 0.3$) can only be explained by the presence of star spots if they are cooler than about 3500\,K and cover at least half of the visible stellar surface. The latter is not obvious for a star with only a low modulation amplitude in Ca\,{\sc ii} H\&K and a rotation period of 35.7~d (Boro-Saikia et al. 2016) unless the cool spots are evenly distributed across stellar longitudes and remain like that.

\begin{figure}[htb]
\centering
\mbox{\includegraphics[angle=0,width=86mm,trim=20 30 30 10]{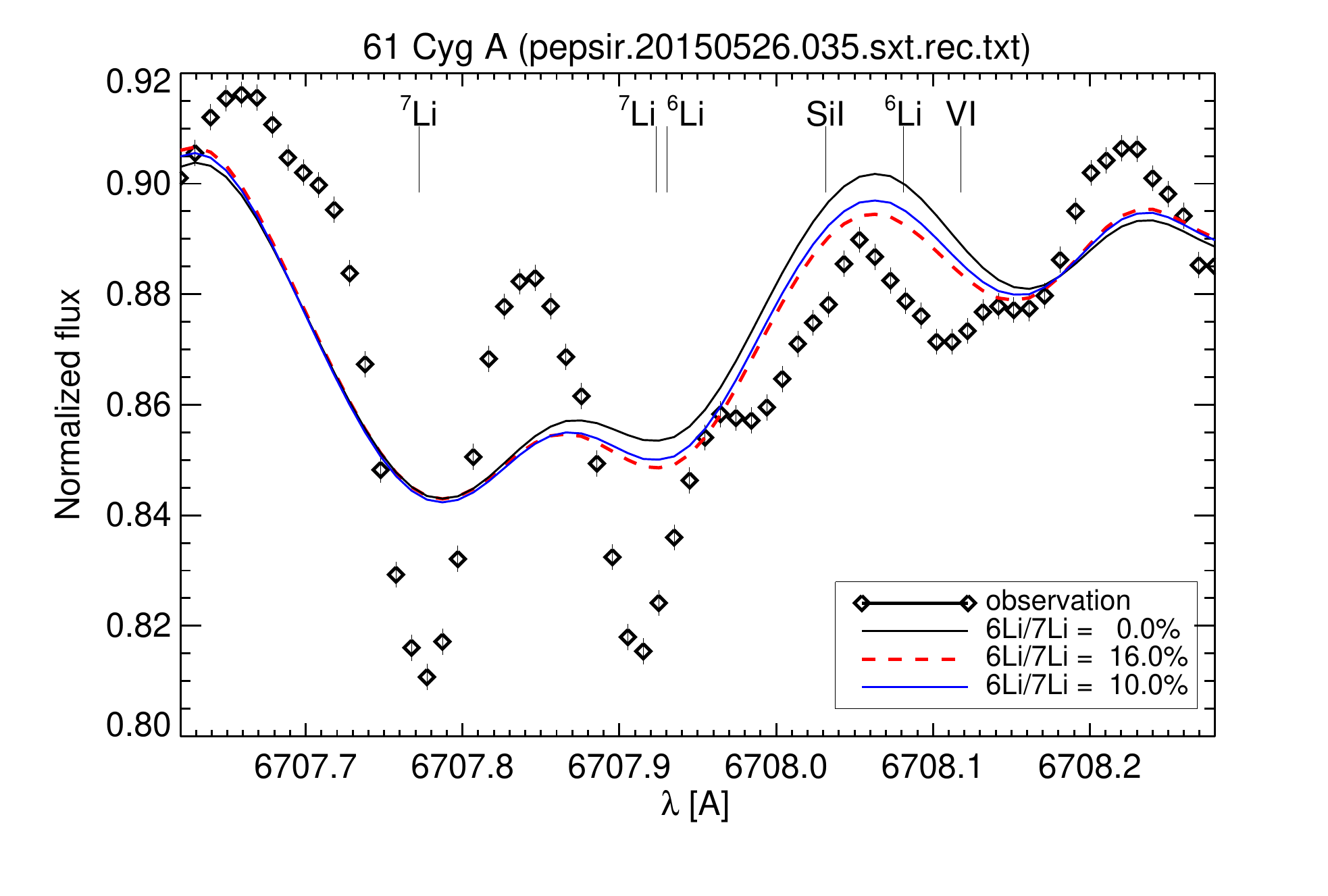}}
\caption{1D-LTE fits for 61\,Cyg\,A for three isotope ratios (lines as labeled) and a spot model with $T_{\rm spot}$ = 3500\,K, and a filling factor $a$ of 0.5. Otherwise as in Fig.~\ref{F4}b.}\label{FA2}
\end{figure}

Figure~\ref{FA2} shows the spot-model Li fit for 61\,Cyg\,A. Shown is the fit with \Teff\ = $T_0$ = 4374\,K, a spot temperature of 3500\,K, and a filling factor $a$ of 0.5. The Li abundance from this fit is A(Li) = 0.53\,dex. Figure~\ref{FA2} can be directly compared with the non-spot results in Fig.~\ref{F4}b. A considerable shift in flux scale can be noted ($-0.07$). Table~\ref{TA2} is an excerpt of spot models with two rather extreme spot filling factors of $a=0.5$ and $a=0.6$. We also note that the fits including spots require super-solar metallicity but with a slightly lowered level of [Fe/H]$\approx$0.037\,dex than for the non-spot fit.

\begin{table}[!hb]
\caption{TiO band head at 7055\,\AA\ for star spot model of 61~Cyg~A
  (\Teff=4374\,K)} \label{TA2}
\begin{center}
\begin{tabular}{lcccccc}
\noalign{\smallskip}\hline\noalign{\smallskip}
$a$ & $T_1$ & $T_2$ & $R_\lambda$ & $d_1$ & $d_2$ & $\overline{d}$ \\
\noalign{\smallskip}\hline\noalign{\smallskip}
{            }  & \multicolumn{5}{c}{--- $a=0.5$ ---} \\
\noalign{\smallskip}\hline\noalign{\smallskip}
  0.500 & 4374 & 4374 & 1.000 & 0.068 & 0.068 & 0.068 \\
  0.500 & 4300 & 4444 & 0.858 & 0.108 & 0.043 & 0.073 \\
  0.500 & 4200 & 4529 & 0.704 & 0.188 & 0.023 & 0.091 \\
  0.500 & 4100 & 4604 & 0.584 & 0.302 & 0.013 & 0.119 \\
  0.500 & 4000 & 4671 & 0.488 & 0.450 & 0.008 & 0.153 \\
  0.500 & 3900 & 4730 & 0.411 & 0.624 & 0.005 & 0.185 \\
  0.500 & 3800 & 4783 & 0.349 & 0.814 & 0.003 & 0.213 \\
  0.500 & 3700 & 4830 & 0.298 & 1.009 & 0.002 & 0.233 \\
  0.500 & 3600 & 4873 & 0.258 & 1.209 & 0.001 & 0.249 \\
  0.500 & 3500 & 4911 & 0.226 & 1.427 & 0.001 & 0.264 \\
\noalign{\smallskip}\hline\noalign{\smallskip}
{            }  & \multicolumn{5}{c}{--- $a=0.6$ ---} \\
\noalign{\smallskip}\hline\noalign{\smallskip}
  0.600 & 4374 & 4374 & 1.000 & 0.068 & 0.068 & 0.068 \\
  0.600 & 4300 & 4478 & 0.828 & 0.108 & 0.034 & 0.075 \\
  0.600 & 4200 & 4601 & 0.655 & 0.188 & 0.013 & 0.100 \\
  0.600 & 4100 & 4707 & 0.528 & 0.302 & 0.006 & 0.137 \\
  0.600 & 4000 & 4800 & 0.432 & 0.450 & 0.003 & 0.178 \\
  0.600 & 3900 & 4881 & 0.357 & 0.624 & 0.001 & 0.219 \\
  0.600 & 3800 & 4953 & 0.299 & 0.814 & 0.001 & 0.252 \\
  0.600 & 3700 & 5017 & 0.253 & 1.009 & 0.000 & 0.278 \\
  0.600 & 3600 & 5074 & 0.216 & 1.209 & 0.000 & 0.296 \\
  0.600 & 3500 & 5125 & 0.188 & 1.427 & 0.000 & 0.314 \\
\noalign{\smallskip}\hline\noalign{\smallskip}\\[5mm]
\end{tabular}
\end{center}
\end{table}

\clearpage

\subsubsection{Results for $\xi$~Boo~B}

\begin{figure}{htb}
\centering
\mbox{\includegraphics[angle=0,width=86mm,trim=20 30 30 10]{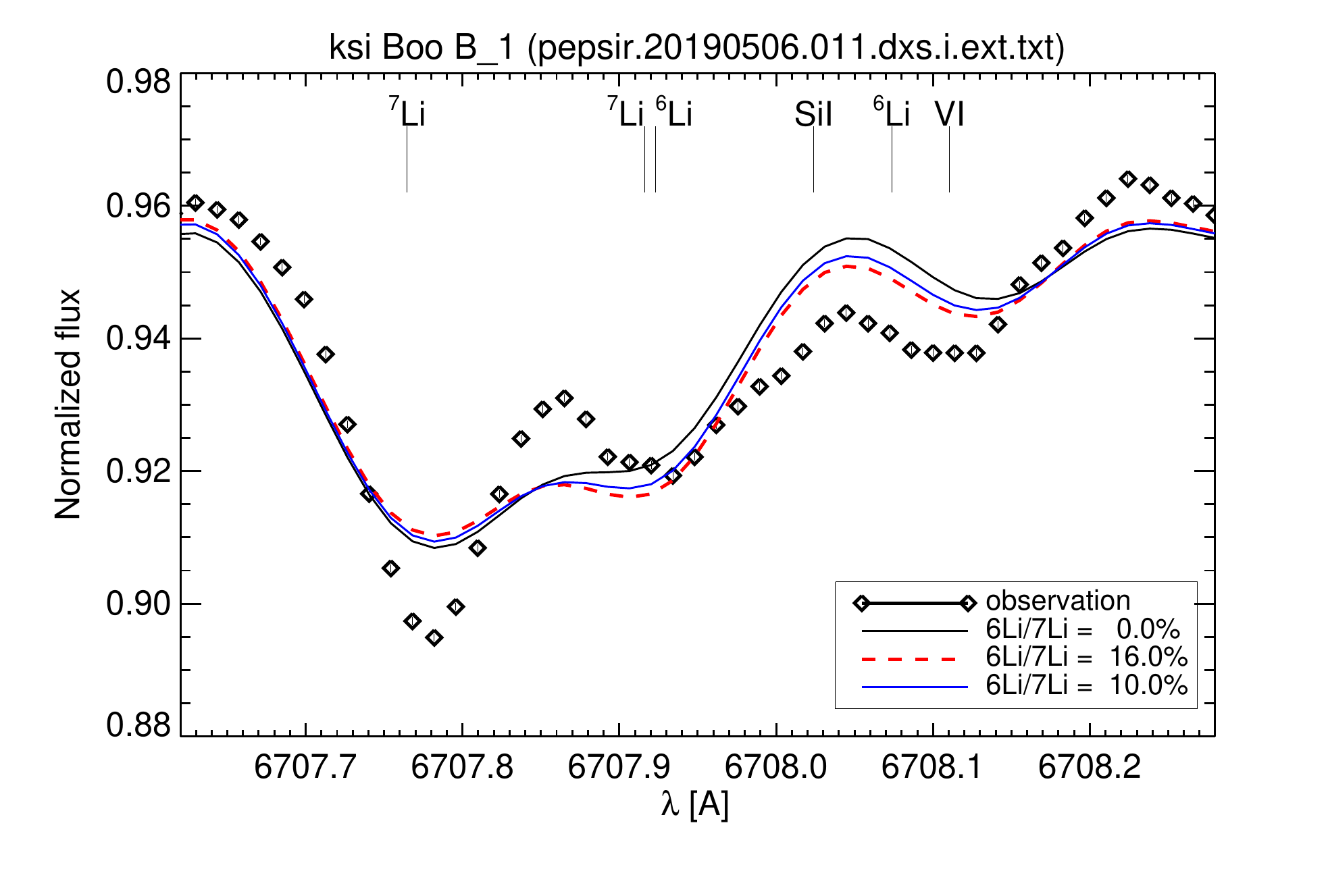}}
\caption{1D-LTE fits for \xiB\ for three isotope ratios (lines as labeled) including a spot model with $T_{\rm spot}$ = 3800\,K, and a filling factor $a$ of 0.3. Otherwise as in Fig.~\ref{F3}b.}\label{FA3}
\end{figure}

Figure~\ref{FA3} shows the spot-model fit for \xiB. Shown is the fit with \Teff\ = $T_0$ = 4570\,K, a spot temperature of 3800\,K, and a filling factor $a$ of 0.3. The Li abundance from this fit is A(Li) = 0.45\,dex. The plot can be directly compared with the non-spot fit in Fig.~\ref{F3}b. Here, the shift in flux scale is $-0.02$. Table~\ref{TA3} is an overview of selected results for \xiB, always assuming an effective temperature of 4570\,K. According to these results, the observed strength of the TiO band head ($\overline{d}\approx 0.1$) can be explained by different star spot configurations. For a filling factor of $0.3$, the spots must have a temperature of about 3800\,K. For larger filling factors of $a=0.4$ or even $0.5$, the necessary spot temperature would be 3900\,K and 4000\,K, respectively. As for 61\,Cyg\,A, we note that the fits including a spot model still require super-solar metallicity but with a slightly lowered level of [Fe/H]$\approx$0.085\,dex.

\begin{table}[!hb]
\caption{TiO band head at 7055\,\AA\ for a star spot model of \xiB\
  (\Teff=4570\,K)}\label{TA3}
\begin{center}
\begin{tabular}{lcccccc}
\noalign{\smallskip}\hline\noalign{\smallskip}
$a$ & $T_1$ & $T_2$ & $R_\lambda$ & $d_1$ & $d_2$ & $\overline{d}$ \\
\noalign{\smallskip}\hline\noalign{\smallskip}
{            }  & \multicolumn{5}{c}{--- $a=0.3$ ---} \\
\noalign{\smallskip}\hline\noalign{\smallskip}
  0.300 &  4570 &  4570 &  1.000 &  0.017 &  0.017 &  0.017 \\
  0.300 &  4500 &  4599 &  0.905 &  0.029 &  0.014 &  0.018 \\
  0.300 &  4400 &  4637 &  0.786 &  0.058 &  0.010 &  0.022 \\
  0.300 &  4300 &  4672 &  0.683 &  0.108 &  0.008 &  0.030 \\
  0.300 &  4200 &  4704 &  0.593 &  0.188 &  0.006 &  0.043 \\
  0.300 &  4100 &  4733 &  0.516 &  0.302 &  0.005 &  0.058 \\
  0.300 &  4000 &  4760 &  0.448 &  0.450 &  0.004 &  0.076 \\
  0.300 &  3900 &  4784 &  0.391 &  0.624 &  0.003 &  0.092 \\
  0.300 &  3800 &  4806 &  0.341 &  0.814 &  0.002 &  0.106 \\
\noalign{\smallskip}\hline\noalign{\smallskip}
{            }  & \multicolumn{5}{c}{--- $a=0.4$ ---} \\
\noalign{\smallskip}\hline\noalign{\smallskip}
  0.400 &  4570 &  4570 &  1.000 &  0.017 &  0.017 &  0.017 \\
  0.400 &  4500 &  4614 &  0.891 &  0.029 &  0.012 &  0.018 \\
  0.400 &  4400 &  4673 &  0.759 &  0.058 &  0.007 &  0.024 \\
  0.400 &  4300 &  4726 &  0.648 &  0.108 &  0.005 &  0.036 \\
  0.400 &  4200 &  4774 &  0.555 &  0.188 &  0.003 &  0.053 \\
  0.400 &  4100 &  4817 &  0.477 &  0.302 &  0.002 &  0.075 \\
  0.400 &  4000 &  4856 &  0.410 &  0.450 &  0.002 &  0.098 \\
  0.400 &  3900 &  4892 &  0.354 &  0.624 &  0.001 &  0.120 \\
  0.400 &  3800 &  4924 &  0.307 &  0.814 &  0.001 &  0.139 \\

\noalign{\smallskip}\hline\noalign{\smallskip}
{            }  & \multicolumn{5}{c}{--- $a=0.5$ ---} \\
\noalign{\smallskip}\hline\noalign{\smallskip}
  0.500 &  4570 &  4570 &  1.000 &  0.017 &  0.017 &  0.017 \\
  0.500 &  4500 &  4636 &  0.872 &  0.029 &  0.010 &  0.019 \\
  0.500 &  4400 &  4722 &  0.724 &  0.058 &  0.005 &  0.027 \\
  0.500 &  4300 &  4799 &  0.606 &  0.108 &  0.003 &  0.042 \\
  0.500 &  4200 &  4867 &  0.510 &  0.188 &  0.001 &  0.064 \\
  0.500 &  4100 &  4928 &  0.432 &  0.302 &  0.001 &  0.092 \\
  0.500 &  4000 &  4982 &  0.367 &  0.450 &  0.001 &  0.121 \\
  0.500 &  3900 &  5031 &  0.314 &  0.624 &  0.000 &  0.149 \\
  0.500 &  3800 &  5075 &  0.269 &  0.814 &  0.000 &  0.173 \\
\noalign{\smallskip}\hline\noalign{\smallskip}
\end{tabular}
\end{center}
\end{table}

\subsubsection{Results for $\xi$~Boo~A}

In this case, we focus only on the strong TiO line at $\lambda\,6708.4$\,\AA\ in the lithium region which, like the 7055\,\AA\ band head, is not detected in our observed spectra of \xiA. This constrains the spot properties of this star. The results of our simple star spot model for \Teff\ = $T_0$ = 5480\,K indicate that spots must not be cooler than about 4000\,K if the filling factor is in the range 0.1 \ldots 0.2. Even larger filling factors, $a > 0.25$, require $T_1 > 4200$\,K. Otherwise this TiO line would be visible in the high-resolution PEPSI spectrum with a line depth of about 1\%, which is not the case.


\begin{thebibliography}{99}
\bibitem[(1981)]{}
Abt, H. A. 1981, ApJS, 45, 437
\bibitem[(2005)]{}
Affer, L., Micela, G., Morel, T., Sanz-Forcada, J., \& Favata, F. 2005, A\&A, 433, 647
\bibitem[(2004)]{}
Allende Prieto, C., Barklem, P. S., Lambert, D. L., \& Cunha, K. 2004, A\&A, 420, 183
\bibitem[(1995)]{}
Anthony-Twarog, B. J., Deliyannis, C. P., Harmer, D., Lee-Brown, D. B., Steinhauer, A., Sun, Q., \& Twarog, B. A. 2018, AJ, 156, 37
\bibitem[(1995)]{}
Baliunas, S. L., Donahue, R. A., Soon, W. H., et al. 1995, ApJ, 438, 269
\bibitem[(1998)]{}
Barklem, P.~S., Anstee, S. D., \& O'Mara, B. J. 1998, PASA, 15, 336
\bibitem[Barklem et al.(2000)]{barklem00}
Barklem, P.~S., Piskunov, N., \& O'Mara, B.~J.\ 2000, \aaps, 142, 467
\bibitem[(2007)]{}
Barnes, S. A. 2007, ApJ, 669, 1167
\bibitem[(2001)]{}
Barrado~y~Navascu{\'e}s, D., Deliyannis, C.~P., \& Stauffer, J.~R.\ 2001, ApJ, 549, 452
\bibitem[(2016)]{}
Boro Saikia, S., Jeffers, S., Morin, J., et al. 2016, A\&A, 594, A29
\bibitem[(2018)]{}
Bouvier, J., Barrado, D., Moraux, E., et al. 2018, A\&A, 613, A63
\bibitem[(2014)]{}
Brooke, J. S. A., Ram, R. S., Western, C. M., Li, G., Schwenke, D. W., \& Bernath, P. F. 2014, ApJS, 210, 23
\bibitem[(1992)]{}
Cayrel de Strobel, G., Hauck, B., Francois, P., Thevenin, F., Friel, E., Mermilliod, M., \& Borde S. 1992, A\&AS, 95, 273
\bibitem[(2017)]{}
Cummings, J.~D., Deliyannis, C.~P., Maderak, R.~M., et al. 2017, AJ, 153, 128.
\bibitem[(2009)]{}
Demory, B.-O., S\'egransan, D., Forveille, T., et al. 2009, A\&A, 505, 205
\bibitem[(1996)]{}
Donahue, R. A., Saar, S. H., \& Baliunas, S. L. 1996, ApJ, 466, 384
\bibitem[(1998)]{}
Fernandes, J., Lebreton, Y., Baglin, A., \& Morel, P. 1998, A\&A, 338, 455
\bibitem[(2022)]{gaia-DR3}
Gaia Collaboration, Vallenari, A., Brown, A. G. A., Prusti, T., et al. 2022, 
A\&A, DOI: 10.1051/0004-6361/202243940
\bibitem[(1994)]{}
Gray, D. F. 1994, PASP, 106, 1248
\bibitem[(2008)]{}
Gustafsson, B., Edvardsson, B., Eriksson, K., Jorgensen, U.G., Nordlund, A., \& Plez, B. 2008, A\&A, 486, 951
\bibitem[(2018)]{}
Harutyunyan, G., Steffen, M., Mott, A., Caffau, E., Israelian, G., Gonz\'alez Hern\'andez, J. I., \& Strassmeier, K. G. 2018, A\&A, 618, A16
\bibitem[(2015)]{}
Heiter, U., Jofr\'e, P., Gustafsson, B., Korn, A. J., Soubiran, C., \& Th\'evenin, F. 2015, A\&A, 582, A49
\bibitem[(2003)]{}
Heiter, U., \& Luck, R. E. 2003, AJ, 126, 2015
\bibitem[(1965)]{}
Herbig, G. H. 1965, ApJ, 141, 588
\bibitem[(2012)]{}
Hill, J. M., Green, R. F., Ashby, D. S., et al. 2012, SPIE, 8444-1
\bibitem[(1974)]{}
Hultqvist, L. 1974, Solar Phys., 34, 25
\bibitem[(1965)]{}
Iben, Jr., I. 1965, ApJ, 141, 993
\bibitem[(1967)]{}
Iben, Jr., I. 1967, ApJ, 147, 650
\bibitem[(2000)]{4A}
Ilyin, I. 2000, PhD Thesis, Univ. of Oulu, Finland
\bibitem[(2015)]{}
Johnstone, C. P., \& G\"udel, M. 2015, A\&A, 578, A129
\bibitem[(1990)]{}
Jorgensen, U.~G. \& Larsson, M.\ 1990, A\&A, 238, 424
\bibitem[(2008)]{}
Kervella, P., M\'erand, A., Pichon, B., et al. 2008, A\&A, 488, 667
\bibitem[(1980)]{}
Kotlar, A.~J., Field, R.~W., Steinfeld, J.~I., et al.\ 1980, Journal of Molecular Spectroscopy, 80, 86
\bibitem[{{Kurucz}(1995)}]{kurucz95}
Kurucz, R.~L. 1995, \apj, 452, 102
\bibitem[{{Kurucz}(2006)}]{kurucz06}
Kurucz, R.~L. 2006, \url{http://kurucz.harvard.edu/atoms/0300/lidlines.dat}
\bibitem[(2014)]{}
Lawler, J. E., Wood, M. P., Den Hartog, E. A., et al. 2014, ApJS, 215, 20
\bibitem[(1978)]{}
Levato H., \& Abt H. A. 1978, PASP, 90, 429
\bibitem[(2009)]{}
Lind, K., Primas, F., Charbonnel, C., Grundahl, F., \& Asplund, M. 2009, A\&A, 503, 545
\bibitem[(1983)]{}
Linsky, J. L., \& Gary, D. E. 1983, ApJ, 274, 776
\bibitem[(2007)]{}
Lockwood G. W., Skiff B. A., Henry G. W., et al. 2007, ApJS, 171, 260
\bibitem[(2017)]{}
Luck, R. E. 2017, AJ, 153, 21
\bibitem[(2005)]{}
Luck, R. E., \& Heiter, U. 2005, AJ, 129, 1063
\bibitem[(2004)]{}
Mandell, A.~M., Ge, J., \& Murray, N.\ 2004, \aj, 127, 1147
\bibitem[(2009)]{mpfit}
Markwardt, C. B. 2009, in Astronomical Data Analysis Software and Systems XVIII, D. Bohlender, P. Dowler \& D. Durand (eds.), ASP Conference Series, Vol. 411, 251
\bibitem[(2019)]{}
McKemmish, L. K., Masseron, T., Hoeijmakers, H. J., Perez-Mesa, V., Grimm, S. L., Yurchenko, S. N., \& Tennyson, J. 2019, MNRAS, 488, 2836
\bibitem[(1999)]{}
Mel{\'e}ndez, J. \& Barbuy, B.\ 1999, \apjs, 124, 527
\bibitem[(2007)]{}
Mel{\'e}ndez, J. \& Cohen, J.~G.\ 2007, \apjl, 659, L25
\bibitem[(2008)]{}
Mel{\'e}ndez, J. \& Asplund, M.\ 2008, A\&A, 490, 817
\bibitem[(2012)]{melendez12}
Mel\'endez, J., Bergemann, M., Cohen, J. G., et al. 2012, A\&A, 543, A29
\bibitem[(2012)]{}
Morgenthaler, A., Petit, P., Saar, S. H., et al. 2012, A\&A, 540, A138
\bibitem[(2003)]{}
Morton, D.~C.\ 2003, \apjs, 149, 205
\bibitem[(2017)]{}
Mott, A., Steffen, M., Caffau, E., Spada, F., \& Strassmeier, K. G. 2017, A\&A, 604, A44
\bibitem[(2020)]{}
Mott, A., Steffen, M., Caffau, E., \& Strassmeier, K. G. 2020, A\&A, 638, A58
\bibitem[(2004)]{}
O'Neal, D., Neff, J. E., Saar, S. H. 1998, ApJ, 507, 919
\bibitem[(2004)]{}
O'Neal, D., Neff, J. E., Saar, S. H., \& Cuntz, M. 2004, AJ, 128, 1802
\bibitem[(2000)]{}
Palmeri, P., Quinet, P., Wyart, J.-F., et al.\ 2000, \physscr, 61, 323
\bibitem[(2005)]{}
Petit, P., Donati, J., Auri\'ere, M., et al. 2005, MNRAS, 361, 837
\bibitem[(1998)]{}
Plez, B. 1998, A\&A, 337, 495
\bibitem[(2012)]{}
Plez, B. 2012, Astrophysics Source Code Library, Turbospectrum
\bibitem[(1999)]{}
Ram, R. S., Bernath, P. F., Dulick, M., \& Wallace, L. 1999, ApJS, 122, 331
\bibitem[(1994)]{}
Randich, S., Giampapa, M., \& Pallavicini, R. 1994, A\&A, 283, 893
\bibitem[(2015)]{}
Ryabchikova, T., Piskunov, N., Kurucz, R. L., Stempels, H. C., Heiter, U., Pakhomov, Yu, \& Barklem, P. S. 2015, Physica Scripta, 90, 054005
\bibitem[(1995)]{}
Ruck, M. J., \& Smith, G. 1995, A\&A, 304, 449
\bibitem[(1992)]{}
Savanov, I. 1992, Bull. of the Crimean Astrophysical Observatory, Vol. 84, p.21
\bibitem[(2015)]{}
Steffen, M., Prakapavicius, D., Caffau, E., et al. 2015, A\&A, 583, A57
\bibitem[(2015)]{pepsi}
Strassmeier, K. G., Ilyin, I., J\"arvinen, A., et al. 2015, AN, 336, 324
\bibitem[(2018a)]{}
Strassmeier, K. G., Ilyin, I., \& Steffen, M. 2018a, A\&A, 612, A44
\bibitem[(2018b)]{}
Strassmeier, K. G., Ilyin, I., \& Weber, M. 2018b, A\&A, 612, A45
\bibitem[(2007)]{}
Takeda, G., Ford, E. B., Sills, A., Rasio, F. A., Fischer, D. A., \& Valenti J. A. 2007, ApJS, 168, 297
\bibitem[(1988)]{}
Toner, C. G., \& Gray, D. F. 1988, ApJ, 334, 1008
\bibitem[(1991)]{}
Toner, C. G., \& LaBonte, B. J. 1991, ApJ, 368, 633
\bibitem[(2005)]{}
Valenti, J. A., \& Fischer, D. A. 2005, ApJS, 159, 141
\bibitem[(1981)]{}
Vogt, S. S. 1981, ApJ, 247, 975
\bibitem[(1999)]{}
Wallace, L., Livingston, W., Bernath, P. F., \& Ram, R. S. 1999, An Atlas of the Sunspot Umbral Spectrum in the Red and Infrared from 8900 to 15,050 cm$^{-1}$ Revised, NSO Techn. Report \#99-001
\bibitem[(1962)]{}
Wielen, R. 1962, AJ, 67, 599
\bibitem[(1963)]{}
Wilson, O. C. 1963, PASP, 75, 62
\bibitem[(1978)]{}
Wilson, O. C. 1978, ApJ, 226, 379
\bibitem[(2018)]{}
Wood, B. E., Laming, J. M., Warren, H. P., \& Poppenhaeger, K. 2018, ApJ, 862, 66
\bibitem[(2004)]{}
Wright, J. T., Marcy, G. W., Butler, R. P., \& Vogt, S. S. 2004, ApJS, 152, 261
\bibitem[(2003)]{}
Xu, H.~L., Svanberg, S., Quinet, P., et al.\ 2003, Journal of Physics B Atomic Molecular Physics, 36, 4773

\end{thebibliography}
\end{document}